\documentclass[format=acmsmall, review=false, screen=true]{acmart}

\usepackage{booktabs} 

\usepackage{bbm}
\usepackage{verbatim}
\usepackage{array}

\usepackage[linesnumbered,ruled,vlined,lined,boxed,commentsnumbered]{algorithm2e}
\usepackage{amsmath}
\usepackage{tikz}
\usetikzlibrary{positioning}
\usetikzlibrary{matrix,shapes,arrows,positioning,chains, calc}
\usepackage{subfigure}
\usepackage{makecell}

\SetAlFnt{\small}
\SetAlCapFnt{\small}
\SetAlCapNameFnt{\small}
\SetAlCapHSkip{0pt}
\IncMargin{-\parindent}

\acmJournal{TOIS}
\acmVolume{9}
\acmNumber{4}
\acmArticle{39}
\acmYear{2017}
\acmMonth{3}
\copyrightyear{2009}

\setcopyright{acmlicensed}

\acmDOI{0000001.0000001}

\received{September 2016}
\received[revised]{December 2017}
\received[accepted]{March 2017}

\begin{document}
\title{Yum-me: A Personalized Nutrient-based Meal Recommender System}  
\author{Longqi Yang}
\affiliation{%
  \institution{Cornell Tech, Cornell University}
  \city{New York City}
  \state{NY}
  \postcode{10011}
  \country{USA}}

\author{Cheng-Kang Hsieh}
\affiliation{%
  \institution{University of California, Los Angeles}
  \city{Los Angeles}
  \state{CA}
  \postcode{90095}
  \country{USA}}

\author{Hongjian Yang}
\affiliation{%
  \institution{Cornell Tech, Cornell University}
  \city{New York City}
  \state{NY}
  \postcode{10011}
  \country{USA}}

\author{John P. Pollak}
\affiliation{%
  \institution{Cornell Tech, Cornell University}
  \city{New York City}
  \state{NY}
  \postcode{10011}
  \country{USA}}

\author{Nicola Dell}
\affiliation{%
  \institution{Cornell Tech, The Jacobs Institute, Cornell University}
  \city{New York City}
  \state{NY}
  \postcode{10011}
  \country{USA}}

\author{Serge Belongie}
\affiliation{%
  \institution{Cornell Tech, Cornell University}
  \city{New York City}
  \state{NY}
  \postcode{10011}
  \country{USA}}

\author{Curtis Cole}
\affiliation{%
  \institution{Weill Cornell Medical College, Cornell University}
  \city{New York City}
  \state{NY}
  \postcode{10011}
  \country{USA}}

\author{Deborah Estrin}
\affiliation{%
  \institution{Cornell Tech, Cornell University}
  \city{New York City}
  \state{NY}
  \postcode{10011}
  \country{USA}}

\begin{abstract}
Nutrient-based meal recommendations have the potential to help individuals prevent or manage conditions such as diabetes and obesity. However, learning people's food preferences and making recommendations that simultaneously appeal to their palate and satisfy nutritional expectations are challenging. Existing approaches either only learn high-level preferences or require a prolonged learning period. We propose \textbf{Yum-me}, a personalized nutrient-based meal recommender system designed to meet individuals' nutritional expectations, dietary restrictions, and fine-grained food preferences. Yum-me enables a simple and accurate food preference profiling procedure via a visual quiz-based user interface, and projects the learned profile into the domain of nutritionally appropriate food options to find ones that will appeal to the user. We present the design and implementation of Yum-me, and further describe and evaluate two innovative contributions. The first contriution is an open source state-of-the-art food image analysis model, named \textbf{FoodDist}. We demonstrate FoodDist's superior performance through careful benchmarking and discuss its applicability across a wide array of dietary applications. The second  contribution is a novel online learning framework that learns food preference from item-wise and pairwise image comparisons. We evaluate the framework in a field study of 227 anonymous users and demonstrate that it outperforms other baselines by a significant margin. We further conducted an end-to-end validation of the feasibility and effectiveness of Yum-me through a 60-person user study, in which Yum-me improves the recommendation acceptance rate by 42.63\%.
\end{abstract}

%
%

\ccsdesc[500]{Information systems~Information retrieval}
\ccsdesc[500]{Information systems~Users and interactive retrieval}
\ccsdesc[500]{Information systems~Personalization}

%
%


\keywords{Nutrient-based meal recommendation; personalization; visual interface; food preferences, online learning}

\thanks{This work is funded through Awards from NSF (\#1344587, \#1343058) and NIH (\#1U54EB020404); as well as gift funding from AOL, RWJF, UnitedHealth Group, Google, and Adobe.

Author's addresses: L. Yang, Department of Computer Science, Cornell Tech, Cornell University; email: ylongqi@cs.cornell.edu; C. Hsieh, Department of Computer Science, UCLA; email: changun@cs.ucla.edu; H. Yang, Cornell Tech, Cornell University; email: hy457@cornell.edu; J. Pollak, Cornell Tech, Cornell University; email: jpp9@cornell.edu; N. Dell, Department of Information Science, Cornell Tech, The Jacobs Institute, Cornell University; email: nixdell@cornell.edu; S. Belongie, Department of Computer Science, Cornell Tech, Cornell University; email: sjb344@cornell.edu; C. Cole, Weill Cornell Medical College, Cornell University; email: ccole@med.cornell.edu; D. Estrin, Department of Computer Science, Cornell Tech, Cornell University; email: destrin@cornell.edu.}

\maketitle

\renewcommand{\shortauthors}{L. Yang et al.}

\section{Introduction}
\label{sec:intro}

Healthy eating plays a critical role in our daily well-being and is indispensable in preventing and managing conditions such as diabetes, high blood pressure, cancer, mental illnesses, and asthma, etc.~\cite{povey2007diabetes,bodnar2005nutrition}. In particular, for children and young people, the adoption of healthy dietary habits has been shown to be beneficial to early cognitive development~\cite{shepherd2006young}. Many applications designed to promote healthy behaviors have been proposed and studied~\cite{kadomura2014persuasive,chang2014lunch,kadomura2013sensing,consolvo2008activity}. Among those applications, the studies and products that target healthy meal recommendations have attracted much attention~\cite{van2011deriving,platejoy}. Fundamentally, the goal of these systems is to suggest food alternatives that cater to individuals' health goals and help users develop healthy eating behavior by following the recommendations~\cite{healthevidence}. Akin to most recommender systems, learning users' preferences is a necessary step in recommending healthy meals that users are more likely to find desirable~\cite{healthevidence}. However, the current food preference elicitation approaches, including 1) on-boarding surveys and 2) food journaling, still suffer from major limitations, as discussed below. 

\begin{itemize}
    \item \textbf{Preferences elicited by surveys are coarse-grained}. A typical on-boarding survey asks a number of multi-choice questions about general food preferences. For example, PlateJoy~\cite{platejoy}, a daily meal planner app, elicits preferences for healthy goals and dietary restrictions with the following questions: 

     \textit{(1) How do you prefer to eat? No restrictions, dairy free, gluten free, kid friendly, pescatarian, paleo, vegetarian...} 

     \textit{(2) Are there any ingredients you prefer to avoid? avocado, eggplant, eggs, seafood, shellfish, lamb, peanuts, tofu....} 

     While the answers to these questions can and should  be used to create a rough dietary plan and avoid clearly unacceptable choices, they do not generate meal recommendations that cater to each person's fine-grained food preferences, and this may contribute to their lower than desired recommendation-acceptance rates, as suggested by our user testing results.
    
    \item \textbf{Food journaling approach suffers from cold-start problem and is hard to maintain.} For example, Nutrino~\cite{nutrino}, a personal meal recommender, asks users to log their daily food consumption and learn users' fine-grained food preferences. As is typical of systems relying on user-generated data, food journaling suffers from the \textit{cold-start} problem, where recommendations cannot be made or are subject to low accuracy when the user has not yet generated a sufficient amount of data. For example, a previous study showed that an active food-journaling user makes about 3.5 entries per day~\cite{cordeiro2015rethinking}. It would take a non-trivial amount of time for the system to acquire sufficient data to make recommendations, and the collected samples may be subject to sampling biases as well~\cite{cordeiro2015rethinking, klesges1995underreports}. Moreover, the photo food journaling of all meals is a habit difficult to adopt and maintain, and therefore is not a generally applicable solution to generate complete food inventories~\cite{cordeiro2015rethinking}.
\end{itemize}
 

To tackle these limitations, we develop \textbf{Yum-me}, a meal recommender that learns \textbf{fine-grained food preferences without relying on the user's dietary history}. We leverage people's apparent desire to engage with food photos\footnote{Collecting, sharing and appreciating high quality, delicious-looking food images is a growing fashion in our everyday lives. For example, food photos are immensely popular on Instagram ( \textit{\#food} has over 177M posts and \textit{\#foodporn} has over 91M posts at the time of writing).} to create a more user-friendly medium for asking visually-based diet-related questions - The recommender learns users' fine-grained food preferences through a simple quiz-based visual interface~\cite{yang2015plateclick} and then attempts to generate meal recommendations that cater to the user's health goals, food restrictions, as well as personal appetite for food. It can be used by people who have food restrictions, such as vegetarian, vegan, kosher, or halal. Particularly, we focus on the health goals in the form of nutritional expectations, e.g. adjusting calories, protein, and fat intake. The mapping from health goals to nutritional expectations can be accomplished by professional nutritionists or personal coaches and is out of the scope of this paper. We leave it as future work. In designing the visual interface~\cite{yang2015plateclick}, we propose a novel online learning framework that is suitable for learning users' potential preferences for a large number of food items while requiring only a modest number of interactions. Our online learning approach balances exploitation-exploration and takes advantage of food similarities through preference-propagation among locally connected graphs. To the best of our knowledge, this is the first interface and algorithm that learns users' food preferences through real-time interactions without requiring specific diet history information. 

For such an online learning algorithm to work, one of the most critical components is a robust food image analysis model. Towards that end, as an additional contribution of this work we present a novel, unified food image analysis model, called \textbf{FoodDist}.
Based on deep convolutional networks and multi-task learning~\cite{krizhevsky2012imagenet,bossard2014food}, FoodDist is the best-of-its-kind Euclidean distance embedding for food images, in which similar food items have smaller distances while dissimilar food items have larger distances. FoodDist allows the recommender to learn users' fine-grained food preferences accurately via similarity assessments on food images. Besides preference learning, FoodDist can be applied to other food-image-related tasks, such as food image detection, classification, retrieval, and clustering. We benchmark FoodDist with the Food-101 dataset~\cite{bossard2014food}, the largest dataset for food images. The results suggest the superior performance of FoodDist over prior approaches~\cite{yang2015plateclick,meyers2015im2calories,bossard2014food}. FoodDist will be made available on Github upon publication.

We evaluate our online learning framework in a field study of 227 anonymous users and we show that it is able to predict the food items that a user likes or dislikes with high accuracy. Furthermore, we evaluate the desirability of Yum-me recommendations end-to-end through a 60-person user study, where each user rates the meal recommendations made by Yum-me relative to those made using a traditional survey-based approach. The study results show that, compared to the traditional survey based recommender, our system significantly improves the acceptance rate of the recommended healthy meals by 42.63\%.  We see Yum-me as a complement to the existing food preference elicitation approaches that further filters the food items selected by a traditional onboarding survey based on users' fine-grained taste for food, and allows a system to serve tailored recommendations upon the first use of the system. We discuss some potential use cases in section \ref{sec:discussion}.


The rest of the paper is organized as follows.  After discussing related work in section \ref{sec:related_work}, we introduce the structure of \textbf{Yum-me} and our backend database in section \ref{sec:yum-me}. In section \ref{sec:online_learning}, we describe the algorithmic details of the proposed online learning algorithm, followed by the architecture of \textbf{FoodDist} model in section \ref{sec:fooddist}. The evaluation results of each component, as well as the recommender are presented in section \ref{sec:evaluation}. Finally, we discuss the limitations, potential impact and real world applications in section \ref{sec:discussion} and conclude in section 
\ref{sec:conclusion}.

\section{Related Work}
\label{sec:related_work}

Our work benefits from, and is relevant to, multiple research threads: (1) healthy meal recommender system, (2) cold-start problem and preference elicitation, (3) pairwise algorithms for recommendation, and (4) food image analysis, which will be surveyed in detail next.

\subsection{Healthy meal recommender system}

Traditional food and recipe recommender systems learn users' dietary preferences from their online activities, including ratings~\cite{forbes2011content,freyne2010intelligent,harvey2013you,elsweiler2015towards}, past recipe choices~\cite{svensson2005designing,geleijnse2011personalized}, and browsing history~\cite{ueda2014recipe,van2011deriving,nutrino}. For example,~\cite{svensson2005designing} builds a social navigation system that recommends recipes based on the previous choices made by the user;~\cite{van2011deriving} proposes to learn a recipe similarity measure from crowd card-sorting and make recommendations based on the self-reported meals; and~\cite{harvey2013you,elsweiler2015towards} generates healthy meal plans based on user's ratings towards a set of recipes and the nutritional requirements calculated for the persona. In addition, previous recommenders also seek to incorporate users' food consumption histories recorded by the food logging and journaling systems (e.g. taking food images~\cite{cordeiro2015rethinking} or writing down ingredients and meta-information~\cite{van2011deriving}).

The above systems, while able to learn users' detailed food preference, share a common limitation, that is they need to wait until a user generates enough data before their recommendations can be effective for this user (i.e., the cold-start problem). Therefore, most commercial applications, for example, Zipongo~\cite{zipongo} and Shopwell~\cite{shopwell} adopt onboarding surveys to more quickly elicit users' \textbf{coarse-grained} food preferences. For instance, Zipongo's questionnaires~\cite{zipongo} ask users about their nutrient intake, lifestyle, habits, and food preferences, and then make day-to-day and week-to-week healthy meals recommendations; ShopWell's survey~\cite{shopwell} are designed to avoid certain food allergens, e.g., gluten, fish, corn, or poultry, and find meals that match to particular lifestyles, e.g., healthy pregnancy or athletic training.

Yum-me fills a vacuum that the prior approaches were not able to achieve, namely a rapid elicitation of users' \textbf{fine-grained} food preferences for immediate healthy meal recommendations. Based on the online learning framework~\cite{yang2015plateclick}, Yum-me infers users' preferences for each single food item among a large food dataset, and projects these preferences for general food items into the domain that meets each individual user's health goals. 

\subsection{Cold-start problem and preference elicitation}

To alleviate the cold-start problem mentioned above, several models of preference elicitation have been proposed in recent years. The most prevalent method of elicitation is to train decision trees to poll users in a structured fashion~\cite{rashid2002getting,golbandi2011adaptive,zhou2011functional,das2013learning,sun2013learning}. These questions are either generated in advance and remain static~\cite{rashid2002getting} or change dynamically based on real-time user feedback~\cite{golbandi2011adaptive,zhou2011functional,das2013learning,sun2013learning}. Also, another previous work explores the possibility of eliciting item ratings directly from the user~\cite{zhang2015dualds,chang2015cscw}. This process can either be carried at item-level~\cite{zhang2015dualds} or within-category (e.g., movies)~\cite{chang2015cscw}.

The preference elicitation methods we mentioned above largely focus on the domain of movie recommendations~\cite{sun2013learning,rashid2002getting,chang2015cscw,zhang2015dualds} and visual commerce~\cite{das2013learning} (e.g., cars, cameras) where items can be categorized based on readily available metadata. 
When it comes to real dishes, however, categorical data (e.g., cuisines) and other associated information (e.g., cooking time) possess a much weaker connection to a user's food preferences. Therefore, in this work, we leverage the visual representation of each meal so as to better capture the process through which people make diet decisions.

\subsection{Pairwise algorithms for recommendation}
Pairwise approaches~\cite{rendle2014improving,park2009pairwise,rendle2009bpr,hsieh2017cml,yang2017personalizing,weston2010large,weston2013learning} are widely studied in recommender system literature. For example, Bayesian Personalized Ranking (BPR)~\cite{rendle2009bpr,rendle2014improving} and Weighted
Approximate-Rank Pairwise (WARP) loss~\cite{weston2010large}, which learn users' and items' representations from user-item pairs, are two representative and popular approaches under this category. Such algorithms have successfully powered many state-of-the-art systems~\cite{hsieh2017cml,weston2013learning}. In terms of the cold-start scenario,~\cite{park2009pairwise} developed a pairwise method to leverage users' demographic information in recommending new items. 

Compared to previous methods, our problem setting is fundamentally different in the sense that \textbf{Yum-me} elicits preferences in an active manner where the input is incremental and contingent on the previous decisions made by the algorithm, while prior work focuses on the static circumstances where the training data is available up-front, and there is no need for the system to actively interact with the user.

\subsection{Food image analysis}

The tasks of analyzing food images are very important in many ubiquitous dietary applications that actively or passively collect food images from mobile~\cite{cordeiro2015rethinking} and wearable~\cite{arab2011feasibility,thomaz2013feasibility},~\cite{ng2015understanding} devices. The estimation of food intake and its nutritional information is helpful to our health~\cite{noronha2011platemate} as it provides detailed records of our dietary history. Previous work mainly conducted the analysis by leveraging the crowd~\cite{noronha2011platemate,turner2015use} and computer vision algorithms~\cite{bossard2014food,meyers2015im2calories}.

Noronha et al.~\cite{noronha2011platemate} crowdsourced nutritional
analysis of food images by leveraging the wisdom of untrained crowds. The study demonstrated the possibility of estimating a meal's calories, fat, carbohydrates, and protein by aggregating the opinions from a large number of people;~\cite{turner2015use} elicit the crowd to rank the healthiness of several food items and validate the results against the ground truth provided by trained observers. Although this approach has been justified to be accurate, it inherently requires human resources that restrict it from scaling to large number of users and providing real time feedback.

To overcome the limitations of crowds and automate the analysis process, numerous papers discussing algorithms for food image analysis, including classification~\cite{bossard2014food,meyers2015im2calories,kawano2014food,beijbom2015menu}, retrieval~\cite{kitamura2009foodlog}, and nutrient estimation~\cite{meyers2015im2calories,sudo2014estimating,chae2011volume,he2013food}. Most of the previous work~\cite{bossard2014food} leveraged hand-crafted image features. However, traditional approaches were only demonstrated in special contexts, such as in a specific restaurant~\cite{beijbom2015menu} or for particular type of cuisine~\cite{kawano2014food} and the performance of the models might degrade when they are applied to food images in the wild.

In this paper, we designed FoodDist using deep convolutional neural network based multitask learning~\cite{caruana1997multitask}, which has been shown to be successful in improving model generalization power and performance in several applications~\cite{zhang2014facial,dai2015instance}. The main challenge of multitask learning is to design appropriate network structures and sharing mechanisms across tasks. With our proposed network structure, we show that FoodDist achieves superior performance when applied to the largest available real-world food image dataset~\cite{bossard2014food}, and when compared to prior approaches.

\section{Yum-me: Personalized Nutrient-based Meal Recommendations}
\label{sec:yum-me}

Our personalized nutrient-based meal recommendation system, Yum-me, operates over a given inventory of food items and suggests the items that will appeal to the users' palate and meet their nutritional expectations and dietery restrictions. A high-level overview of Yum-me's recommendation process is shown in Fig.~\ref{fig:plateclick} and briefly described as follows:

\begin{figure}
  \centering
  \includegraphics[width=1\columnwidth]{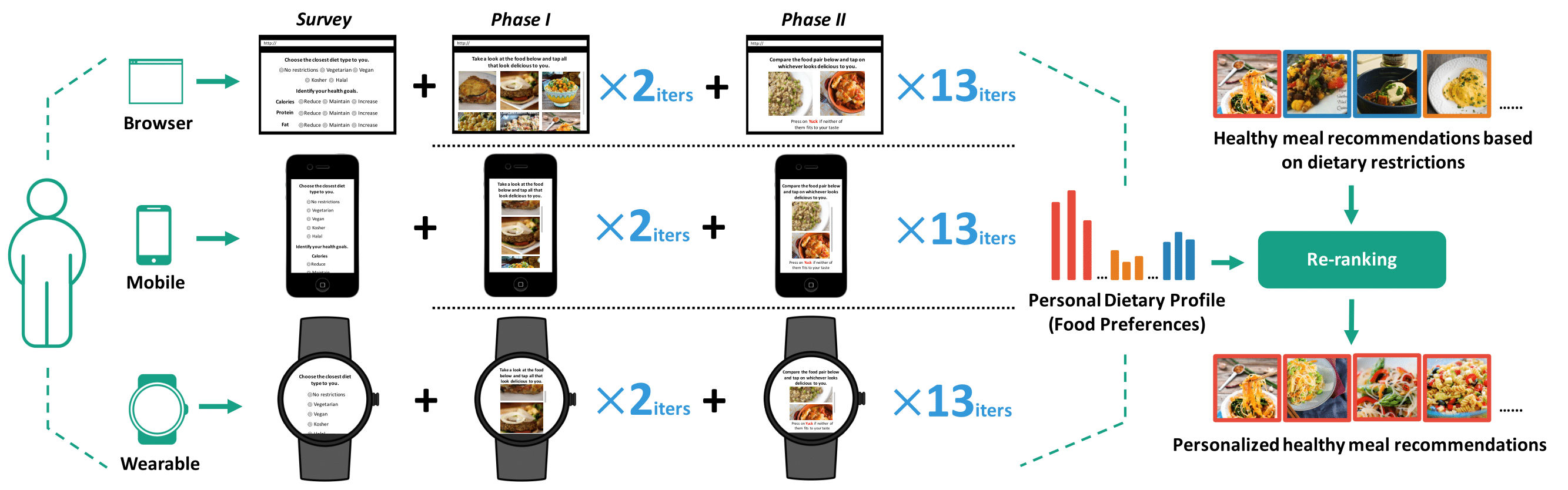}
  \caption{Overview of Yum-me. This figure shows three sample scenarios in which Yum-me can be used: desktop browser, mobile, and smart watch. The fine-grained dietary profile is used to re-rank and personalize meal recommendations.}
    \label{fig:plateclick}
\end{figure}

\begin{itemize}
    \item \textit{Step 1}: Users answer a simple survey to specify their dietary restrictions and nutritional expectations. This information is used by Yum-me to filter food items and create an initial set of recommendation candidates.
    \item \textit{Step 2}: Users then use an adaptive visual interface to express their fine-grained food preferences through simple comparisons of food items. The learned preferences are used to further re-rank the recommendations presented to them.
\end{itemize}

In the rest of this section, we describe our backend large-scale food database and aforementioned two recommendation steps:  1) a user survey that elicits user's dietary restrictions and nutritional expectations, and 2) an adaptive visual interface that elicits users' fine-grained food preferences.

\subsection{Large scale food database}

To account for the dietary restrictions in many cultures and religions, or people's personal choices, we prepare a separate food database for each of the following dietary restrictions:

\begin{center}
\textbf{No restrictions, Vegetarian, Vegan, Kosher, Halal} \footnote{Our system is not restricted to these five dietary restrictions and we will extend the system functionalities to other categories in the future.}
\end{center}

For each diet type, we pulled over 10,000 main dish recipes along with their images and metadata (ingredients, nutrients, tastes, etc.) from the Yummly API~\cite{yummly}. The total number of recipes is around 50,000. In order to customize food recommendations for people with specific dietary restrictions, e.g., vegetarian and vegan, we filter recipes by setting the \textit{allowedDiet} parameter in the search API. For kosher or halal, we explicitly rule out certain ingredients by setting \textit{excludedIngredient} parameter. The lists of excluded ingredients are shown as below:

\begin{itemize}
    \item \textbf{Kosher: }pork, rabbit, horse meat, bear, shellfish, shark, eel, octopus, octopuses, moreton bay bugs, frog.
    \item \textbf{Halal: }pork, blood sausage, blood, blood pudding, alcohol, grain alcohol, pure grain alcohol, ethyl alcohol.
\end{itemize}

One challenge in using a public food image API is that many recipes returned by the API contain non-food images and incomplete nutritional information. Therefore, we further filter the items with the following criteria: the recipe should have 1) nutritional information of calories, protein and fat, and 2) at least one food image. In order to automate this process, we build a binary classifier based on a deep convolutional neural network to filter out non-food images. As suggested by~\cite{meyers2015im2calories}, we treat the whole training set of Food-101 dataset~\cite{bossard2014food} as one generic \textit{food} category and sampled the same number of images (75,750) from the ImageNet dataset~\cite{deng2009imagenet} as our \textit{non-food} category. We took the pretrained VGG CNN model~\cite{Simonyan14c} and replaced the final 1000 dimensional softmax with a single logistic node. For the validation, we use the Food-101 testing dataset along with the same number of images sampled from ImageNet (25,250). We trained the binary classifier using the Caffe framework~\cite{jia2014caffe} and it reached 98.7\% validation accuracy. We applied the criteria to all the datasets and the final statistics are shown in Table.~\ref{tab:size}. 

\begin{figure}
  \centering
  \includegraphics[width=1\columnwidth]{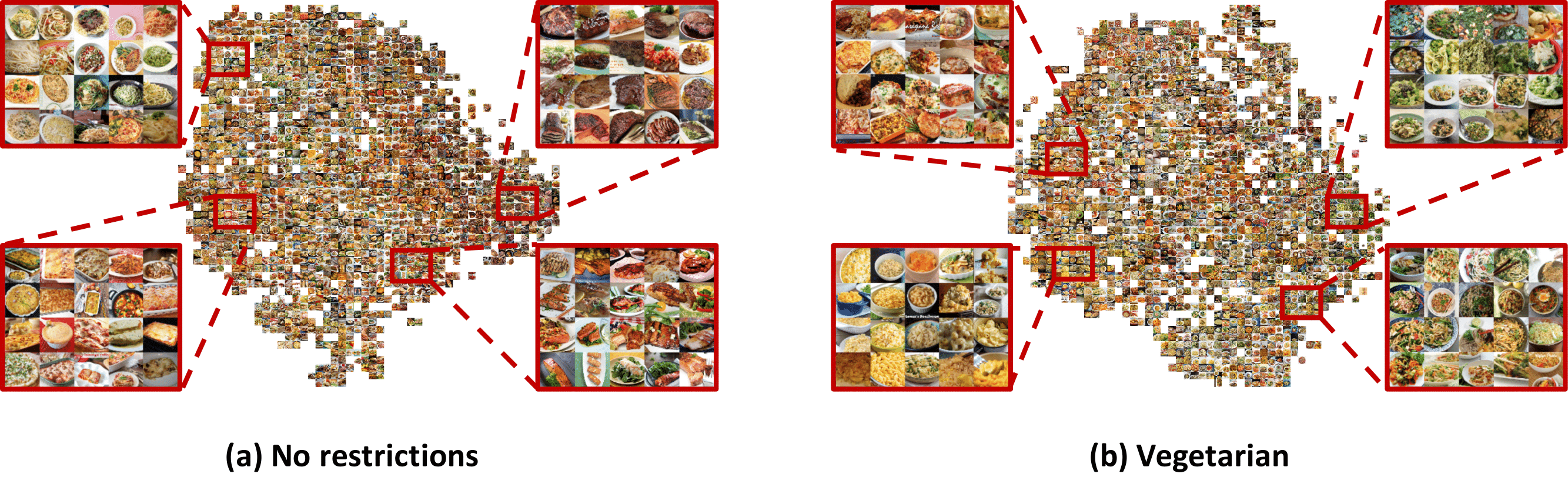}
  \caption{Overview of two sample databases: (a) Database for users without dietary restrictions, (b) Database for vegetarian users. }
    \label{fig:embedding}
\end{figure}

Fig.~\ref{fig:embedding} shows the visualizations of the collected datasets. For each of the recipe images, we embed it into an 1000-dimensional feature space using FoodDist (described later in Section \ref{sec:fooddist}) and then project all the images onto a 2-D plane using t-Distributed Stochastic Neighbor Embedding(t-SNE)~\cite{van2008visualizing}. For visibility, we further divide the 2-D plane into several blocks; from each of which, we sample a representative food image residing in that block to present in the figure.  Fig.~\ref{fig:embedding}  demonstrates the large diversity and coverage of the collected datasets. Also, the embedding results clearly demonstrate the effectiveness of FoodDist in grouping similar food items together while pushing dissimilar items away. This is important to the performance of Yum-me, as discussed in Section \ref{sec:yum-me user study}.


\begin{table}
\caption{Sizes of databases that catered to different diet types. Unit: number of unique recipes.}
\label{tab:size}

 \begin{tabular}{|c | c | c |} 
 
 \hline
 Database & Original size & Final size \\
 \hline\hline
 No restriction & 9405 & 7938 \\ 
 \hline
 Vegetarian & 10000 & 6713 \\
 \hline
 Vegan & 9638 & 6013 \\
 \hline
 Kosher & 10000 & 4825 \\
 \hline
 Halal & 10000 & 5002 \\
 \hline
\end{tabular}
\end{table}

\subsection{User survey}
The user survey is designed to elicit user's high-level dietary restrictions and nutritional expectations. Users can specify their dietary restrictions among the five categories mentioned-above and indicate their nutritional expectations in terms of the desired amount of \textbf{calories}, \textbf{protein} and \textbf{fat}. We choose these nutrients for their high relevance to many common health goals, such as weight control~\cite{epstein1985effect}, sports performance~\cite{brotherhood1984nutrition}, etc. We provide three options for each of these nutrients, including \textbf{reduce}, \textbf{maintain}, and \textbf{increase}. The user's diet type is used to select the appropriate food dataset, and the food items in the dataset are further ranked by their suitability to users' health goals based on the nutritional facts.

To measure the suitability of food items given users' nutritional expectations, we rank the recipes in terms of different nutrients in both \textit{ascending} and \textit{descending} order, such that each recipe is associated with six ranking values, i.e., $r_{\text{calories}, a}$, $r_{\text{calories}, d}$, $r_{\text{protein}, a}$, $r_{\text{protein}, d}$, $r_{\text{fat}, a}$ and $r_{\text{fat}, d}$, where $a$ and $d$ stand for \textit{ascending} and \textit{descending} respectively. The final suitability value for each recipe given the health goal is calculated as follows:

\begin{equation}
    u= \sum_{n\in \mathbb{U}} \alpha_{n, a}r_{n, a} + \sum_{n\in \mathbb{U}} \alpha_{n, d}r_{n, d}
\end{equation}

where $\mathbb{U}=\{\text{calories, protein, fat}\}$. The indicator coefficient $\alpha_{n,a}=1 \iff$ nutrient $n$ is rated as \textit{reduce} and $\alpha_{n,d}=1 \iff$ nutrient $n$ is rated as \textit{increase}. Otherwise $\alpha_{n,a}=0$ and $\alpha_{n,d}=0$. If user's goal is to maintain all nutrients, then all recipes are given equal rankings. Eventually, given a user's responses to the survey, we rank the suitability of all the recipes in the corresponding database and select top-$M$ items (around top 10\%) as the candidate pool of healthy meals for this user. In our initial prototype, we set $M=500$.

\subsection{Adaptive visual interface}\label{sec:plate_click}

 Based on the food suitability ranking, a candidate pool of healthy meals is created. However, not all the meals in this candidate pool will suit the user's palate. Therefore, we design an adaptive visual interface to further identify recipes that cater to the user's taste through eliciting their fine-grained food preferences. We propose to learn users' fine-grained food preferences by presenting users with food images and ask them to choose ones that look delicious.
 
 Formally, the food preference learning task can be defined as: given a large \textbf{target set} of food items $\mathbb{S}$, we represent user's preferences as a distribution over all the possible food items, i.e. $\mathbf{p}=[p_{1}, ..., p_{| \mathbb{S} |}], \sum_{i} p_{i}=1$, where each element $p_{i}$ denotes the user's favorable scale for item $i$. Since the number of items, $| \mathbb{S} |$, is usually quite large and intractable to elicit individually from the user \footnote{The target set is often the whole food database that different applications use. For example, the size of Yummly database can be up to 1-million~\cite{yummly}.}, the approach we take is to adaptively choose a specific and much smaller \textbf{subset} $\mathbb{V}$ to present to the user, and propagate the users' preferences for those items to the rest items based on their visual similarity. Specifically, as Fig.~\ref{fig:plateclick} shows, the preference elicitation process can be divided into two phases:

\textit{Phase I:} In each of the first 2 iterations, we present ten food images and ask users to tap on all the items that look delicious to them.

\textit{Phase II:} In each of the subsequent iterations, we present a pair of food images and ask users to either compare the food pair and tap on the one that looks delicious to them or tap on ``Yuck'' if neither of the items appeal to their taste.

In order to support the preference elicitation process, we design a novel exploration-exploitation online learning algorithm built on a state-of-the-art food image embedding model, which will be discussed in the Section \ref{sec:online_learning} and Section \ref{sec:fooddist} respectively.

\section{Online Learning Framework}
\label{sec:online_learning}

\begin{figure} [t]
\centering
\includegraphics[width=0.5\linewidth]{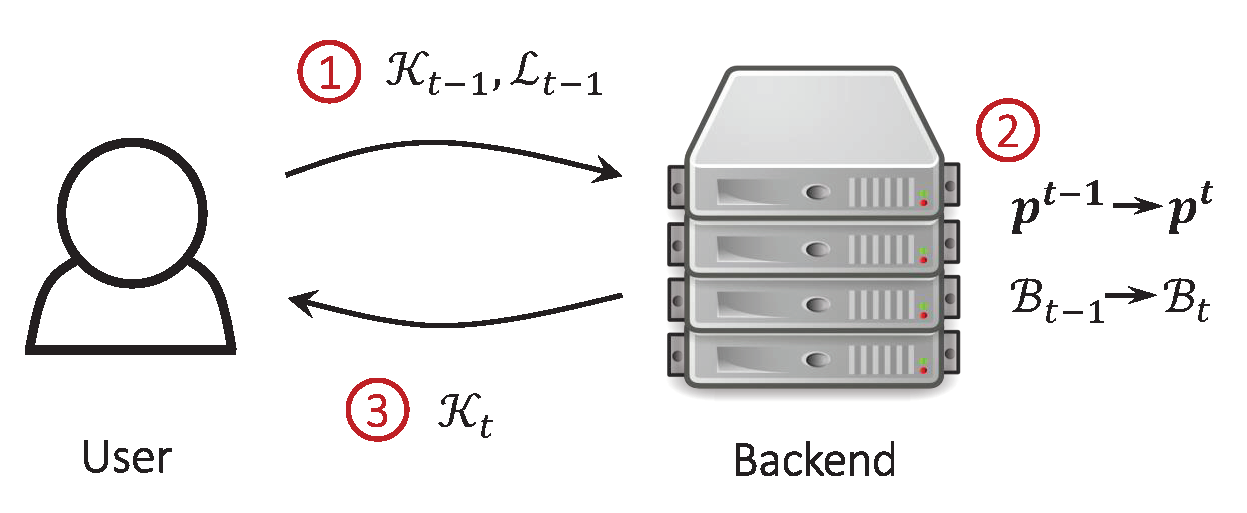}
\caption{User-system interaction at iteration $t$.}
\label{fig:interaction}
\end{figure}

We model the interaction between the user and our backend system at iteration $t, (t \in \mathcal{R}^{+}, t=1,2, ..., T)$ as Fig.~\ref{fig:interaction} shows. The symbols that will be used in our algorithms are defined as follows: 

\begin{itemize}
    \item $\mathcal{K}_{t}:$ Set of food items that are presented to user at iteration $t$ ($\mathcal{K}_{0} = \emptyset$). $\forall k \in \mathcal{K}_t$, $k \in \mathbb{S}$;
    \item $\mathcal{L}_{t-1}:$ Set of food items that user \textit{prefer(select)} among $\{ k|k \in \mathcal{K}_{t-1} \}$. $\mathcal{L}_{t-1} \subseteq \mathcal{K}_{t-1}$;
    \item $\boldsymbol{p}^{t}=[p_{1}^{t}, ..., p_{\mid\mathbb{S}\mid}^{t}]:$ User's preference distribution on all food items at iteration $t$, where $\|\boldsymbol{p}^{t}\|_{1} = 1$. $\boldsymbol{p}^{0}$ is initialized as $p_{i}^{0} = \frac{1}{\mid \mathbb{S} \mid}$;
    \item $\mathcal{B}_{t}:$ Set of food images that have been already explored until iteration $t$ ($\mathcal{B}_{0} = \emptyset$). $\mathcal{B}_{i} \subseteq \mathcal{B}_{j} (i < j)$;
    \item $\mathcal{F}=\{f(x_{1}), ..., f(x_{\mid\mathbb{S}\mid})\}:$ Set of feature vectors of food images $x_{i}(i=1, ..., \mid\mathbb{S}\mid)$ extracted by a feature extractor, denoted by $f$. We use FoodDist as the feature extractor. More details about FoodDist appear in Section \ref{sec:fooddist}.
\end{itemize}

Based on the workflow depicted in Fig.~\ref{fig:interaction}, for each iteration $t$, the backend system updates vector $\boldsymbol{p}^{t-1}$ to $\boldsymbol{p}^{t}$ and set $\mathcal{B}_{t-1}$ to $\mathcal{B}_{t}$ based on users' selections $\mathcal{L}_{t-1}$ and previous image set $\mathcal{K}_{t-1}$. 
After that, it decides the set of images that will be immediately presented to the user (i.e., $\mathcal{K}_{t}$). 
Our food preference elicitation framework can be formalized in Algorithm. \ref{algo:framework}. The core procedures are \textit{update} and \textit{select}, which will be described in the following subsections for more details.

\begin{algorithm}
\SetKwFunction{Select}{select}
\SetKwFunction{Update}{update}
\SetKwFunction{Show}{ShowToUser}
\SetKwFunction{Wait}{WaitForSelection}
\SetKwInOut{Input}{input}
\SetKwInOut{Output}{output}

\KwData{$\mathbb{S}$, $\mathcal{F}=\{f(x_{1}), ..., f(x_{\mid\mathbb{S}\mid})\}$}
\KwResult{$\boldsymbol{p}^{T}$}
\BlankLine
$\mathcal{B}_{0}=\emptyset$,  $\mathcal{K}_{0}=\emptyset$, $\mathcal{L}_{0} = \emptyset$, $\boldsymbol{p}^{0}=[\frac{1}{\mid \mathbb{S} \mid}, ..., \frac{1}{\mid \mathbb{S} \mid}]$ \;
\For{$t\leftarrow 1$ \KwTo $T$}{
$[\mathcal{B}_{t}, \boldsymbol{p}^{t}]\leftarrow$ \Update{$\mathcal{K}_{t-1}$, $\mathcal{L}_{t-1}$, $\mathcal{B}_{t-1}$, $\boldsymbol{p}^{t-1}$} \;
$\mathcal{K}_{t}\leftarrow$ \Select{$t$, $\mathcal{B}_{t}$, $\boldsymbol{p}_{t}$} \;
\uIf{$t$ equals $T$}{return $\boldsymbol{p}^{T}$}
\Else{\Show{$\mathcal{K}_{t}$} \;
$\mathcal{L}_{t}\leftarrow$ \Wait{} \;}
}
\caption{Food Preference Elicitation Framework}
\label{algo:framework}
\end{algorithm}

\subsection{User State Update}
Based on user's selections $\mathcal{L}_{t-1}$ and the image set $\mathcal{K}_{t-1}$, the \textit{update} module renews user's state from $\{\mathcal{B}_{t-1}, \boldsymbol{p}^{t-1}\}$ to $\{\mathcal{B}_{t}, \boldsymbol{p}^{t}\}$. 
Our intuition and assumption behind following algorithm design is that people tend to have close preferences for similar food items. 

\underline{Preference vector $\boldsymbol{p}^{t}$:}
Our strategy of updating preference vector $\boldsymbol{p}^{t}$ is inspired by Exponentiated Gradient Algorithm in bandit settings (EXP3)~\cite{EXP3}. Specifically, at iteration $t$, each $p_{i}^{t}$ in vector $\boldsymbol{p}^{t}$ is updated by:

\begin{equation}
p_{i}^{t} \leftarrow p_{i}^{t-1} \times e^{\frac{\beta u_{i}^{t-1}}{p_{i}^{t-1}}}
\label{eqn:prefer_update}
\end{equation}
where $\beta$ is the exponentiated coefficient that controls update speed and $\boldsymbol{u}^{t-1}=\{u_{1}^{t-1}, ..., u_{\mid \mathbb{S} \mid}^{t-1}\}$ is the update vector used to adjust each preference value.

In order to calculate update vector $\boldsymbol{u}$,
we formalize the user's selection process as a data labeling problem~\cite{zhou2004learning} where for item $i \in \mathcal{L}_{t-1}$, label $y_{i}^{t-1} = 1$ and for item $j \in \mathcal{K}_{t-1} \backslash \mathcal{L}_{t-1}$, label $y_{j}^{t-1} = -1$. Thus, the label vector $\boldsymbol{y}^{t-1}=\{y_{1}^{t-1}, ..., y_{\mid \mathbb{S} \mid}^{t-1}\}$ provided by the user is:

\begin{equation}
y_{i}^{t-1} = \left\{
     \begin{array}{lr}
       1 & : i \in \mathcal{L}_{t-1}\\
       0 & : i \not\in \mathcal{K}_{t-1} \\
       -1 & : i \in \mathcal{K}_{t-1} \backslash \mathcal{L}_{t-1}
     \end{array}
   \right.
\label{eqn:label}
\end{equation}

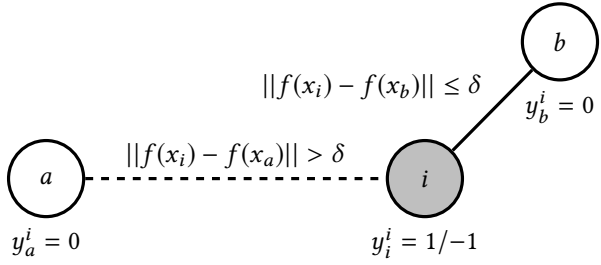
\begin{figure} [t]
\centering
\begin{tikzpicture}
\tikzstyle{mynode}=[minimum width=1cm, very thick, draw, circle];

\node [mynode] (sa) at (0,0) {$a$};
\node [mynode,right = 4cm of sa, fill=gray!50] (si) {$i$};
\node [mynode,above right = 1.5cm of si] (sb) {$b$};

\node [below = 0cm of sa] (aa) {$y_a^i=0$};
\node [below = 0cm of si] (ii) {$y_i^i=1/-1$};
\node [below = 0cm of sb] (bb) {$y_b^i=0$};

\draw[dashed, very thick] (sa) -- (si) node [midway, above] (e_ai) {$||f(x_i) - f(x_a)|| > \delta$};
\draw[very thick] (si) -- (sb) node [midway, above left] (e_ai) {$||f(x_i) - f(x_b)|| \le \delta$};

\end{tikzpicture}
\caption{Locally connected graph with item $i$.}
\label{fig:local}
\end{figure}

For update vector $\boldsymbol{u}$, we expect that it is close to label vector $\boldsymbol{y}$ but with smooth propagation of label values to nearby neighbors (For convenience, we omit superscript that denotes current iteration). 
The update vector $\boldsymbol{u}$ can be regarded as a soften label vector compared with $\boldsymbol{y}$.
To make the solution more computationally tractable, for each item $i$ with $y_{i} \neq 0$, we construct a locally connected undirected graph $G^{i}$ as Fig.~\ref{fig:local} shows: 
$\forall j \in \mathbb{S}$, add an edge $(i, j)$ if $\lVert f(x_{i}) - f(x_{j})\rVert \leq \delta$. 
The labels $\boldsymbol{y}^{i}$ for vertices $s_j$ in graph $G^{i}$ are calculated as $y_{j}^{i}=0 (j=1, \dots, |\mathbb{S}| \setminus i), y_{i}^{i}=y_{i}$. 

For each locally connected graph $G^{i}$, we fix $u_{i}^{i}$ value as $u_{i}^{i}=y_{i}^{i}$ and propose the following regularized optimization method to compute other elements ($\forall u_{j}^{i}, j \neq i$) of update vector $\boldsymbol{u}^{i}$ , which is inspired by the traditional label propagation method~\cite{zhou2004learning}. Consider the problem of minimizing following objective function $Q(\boldsymbol{u}^{i})$:

\begin{equation}
\begin{aligned}
\underset{\boldsymbol{u}^{i}}{\text{min}}
\sum_{j=1, j \neq i}^{\mid \mathbb{S} \mid} w_{ij}(y_{i}^{i} - u_{j}^{i})^{2} + \sum_{j=1, j\neq i}^{\mid \mathbb{S} \mid} (1-w_{ij})(u_{j}^{i} - y_{j}^{i})^{2}
\end{aligned}
\label{eqn:optimization}
\end{equation}

In Eqn. (\ref{eqn:optimization}), $w_{ij}$ represents the similarity measure between food item $s_{i}$ and $s_{j}$:

\begin{equation}
w_{ij} = \left\{
     \begin{array}{lr}
       e^{-\frac{1}{2\alpha^{2}} \lVert f(x_{i}) - f(x_{j}) \rVert^{2}} & : \lVert f(x_{i}) - f(x_{j})\rVert \leq \delta\\
       0 & : \lVert f(x_{i}) - f(x_{j})\rVert > \delta
     \end{array}
   \right.
\end{equation}

where $\alpha^{2} = \frac{1}{\mid \mathbb{S} \mid^{2}} \sum_{i,j \in \mathbb{S}} \lVert f(x_{i}) - f(x_{j}) \rVert^{2}$

The first term of the objective function $Q(\boldsymbol{u}^{i})$ is the \textit{smoothness constraint} as the update value for similar food items should not change too much. The second term is the \textit{fitting constraint}, which makes $\boldsymbol{u}^{i}$ close to the initial labeling assigned by user (i.e. $\boldsymbol{y}^{i}$). However, unlike~\cite{zhou2004learning}, in our algorithm, the trade-off between these two constraints is dynamically adjusted by the similarity between item $i$ and $j$ where similar pairs are weighed more with smoothness and dissimilar pairs are forced to be close to initial labeling.

With Eqn. (\ref{eqn:optimization}) being defined, we can take the partial derivative of $Q(\boldsymbol{u}^{i})$ with respect to different $u_{j}^{i}$ as follows:

\begin{equation}
\frac{\partial Q(\boldsymbol{u}^{i})}{u_{j, j \neq i}^{i}} = 2w_{ij}(u_{j}^{i} - u_{i}^{i}) + 2(1-w_{ij})(u_{j}^{i} - y_{j}^{i}) = 0
\end{equation}

As $u_{i}^{i} = y_{i}^{i}$, then:

\begin{equation}
u_{j}^{i} = w_{ij}u_{i}^{i}=w_{ij}y_{i}^{i} (j=1, 2, ..., \mid \mathbb{S} \mid)
\end{equation}

After all $\boldsymbol{u}^{i}$ are calculated, the original update vector $\boldsymbol{u}$ is then the sum of $\boldsymbol{u}^{i}$, i.e. $\boldsymbol{u} = \sum_{i} \boldsymbol{u}^{i}$. The pseudo code for the algorithm of updating preference vector is shown in Algorithm.\ref{algo:update} for details.

\underline{Explored food image set $\mathcal{B}_{t}$:}
In order to balance the \textit{exploitation} and \textit{exploration} in image selection phase, we maintain a set $\mathcal{B}_{t}$ that keeps track of all similar food items that have already been visited by user and the updating rule for $\mathcal{B}_{t}$ is as follows:

\begin{equation}
\mathcal{B}_{t} \leftarrow \mathcal{B}_{t-1} \cup \{i \in \mathbb{S}|min_{j \in \mathcal{K}_{t-1}} \lVert f(x_{i}) - f(x_{j}) \rVert \leq \delta\}
\end{equation}

With the algorithms designed for updating preference vector $\boldsymbol{p}^{t}$ and explored image set $\mathcal{B}_{t}$, the overall functionality of procedure \textit{update} is shown in Algorithm.\ref{algo:update}.

\begin{algorithm}
\SetKwFunction{Update}{update}
\SetKwFunction{Min}{min}
\SetKwFunction{Normalize}{normalize}
\SetKwInOut{Input}{input}
\SetKwInOut{Output}{output}
\SetKwProg{Fn}{Function}{}{}

\Fn{\Update{$\mathcal{K}_{t-1}, \mathcal{L}_{t-1}, \mathcal{B}_{t-1}, \boldsymbol{p}^{t-1}$}}{
\Input{$\mathcal{K}_{t-1}, \mathcal{L}_{t-1}, \mathcal{B}_{t-1}, \boldsymbol{p}^{t-1}$}
\Output{$\mathcal{B}_{t}, \boldsymbol{p}^{t}$}
\BlankLine
$\boldsymbol{u}=[0, ..., 0], \mathcal{B}_{t}=\mathcal{B}_{t-1}, \boldsymbol{p}^{t}=\boldsymbol{p}^{t-1}$ \\

\For{$i \leftarrow$ 1 \KwTo $\mid \mathbb{S} \mid$}{
    \tcp{preference update}
    \For{$s_{j}$ in $\mathcal{K}_{t-1}$}{
        $u_{i} \leftarrow u_{i}+(-1)^{\mathbbm{1}(j \in \mathcal{L}_{t-1}) - 1}w_{ij}$
    }
    $p_{i}^{t}=p_{i}^{t-1}e^{\frac{\beta u_{i}}{p_{i}^{t-1}}}$ \\
    \tcp{explored image set update}
    \If{\Min{$\lVert f(x_{i}) - f(x_{j}) \rVert$, $\forall j \in \mathcal{K}_{t-1}$} $\leq \delta$}{
        $\mathcal{B}_{t} \leftarrow \mathcal{B}_{t} \cup \{i\}$
    }
}
\tcp{normalize $\boldsymbol{p}^{t}$ s.t.$\|\boldsymbol{p}^{t}\|_{1} = 1$}
\Normalize{$\boldsymbol{p}^{t}$}
}
\caption{User state update Algorithm}
\label{algo:update}
\end{algorithm}

\begin{algorithm}
\SetKwFunction{Percentile}{percentile}
\SetKwFunction{Kmeanspp}{k-means-pp}
\SetKwFunction{Random}{random}
\SetKwInOut{Input}{input}
\SetKwInOut{Output}{output}
\SetKwProg{Fn}{Function}{}{}

\Fn{\Kmeanspp{$\mathbb{S}$, $n$}}{
\Input{$\mathbb{S}$, $n$}
\Output{$\mathcal{K}_{t}$}
\BlankLine
$\mathcal{K}_{t}$=\Random{$\mathbb{S}$} \\
\While{$\mid \mathcal{K}_{t} \mid < n$}{
    prob $\leftarrow [0,...,0]_{\mid \mathbb{S} \mid}$ \\
    \For{$i \leftarrow$ 1 \KwTo $\mid \mathbb{S} \mid$}{
        $\text{prob}_{i} \leftarrow min(\lVert f(x_i) - f(x_j) \rVert^{2} | \forall{j \in \mathcal{K}_{t}})$
    }
    sample $m \in \mathbb{S}$ with probability $\propto \text{prob}_{m}$\\
    $\mathcal{K}_{t} \leftarrow \mathcal{K}_{t} \cup \{m\}$
}
}
\caption{Kmeans++ Algorithm for Exploration}
\label{algo:kmeans}
\end{algorithm}

\subsection{Images Selection}
After updating user state, the \textit{select} module then picks food images that will be presented in the next round. In order to trade-off between exploration and exploitation in our algorithm, we propose different images selection strategies based on current iteration $t$.

\subsubsection{Food Exploration}

For each of the first two iterations, we select ten different food images by using \textit{K-means++}~\cite{arthur2007k} algorithm, which is a seeding method used in \textit{K-means clustering} and can guarantee that selected items are evenly distributed in the feature space. For our use case, \textit{K-means++} algorithm is summarized in Algorithm.\ref{algo:kmeans}.

\subsubsection{Food Exploitation-Exploration}
Starting from the third iteration, users are asked to make pairwise comparisons between food images. To balance the Exploitation and Exploration, we always select one image from the area with higher preference value based on current $\boldsymbol{p}^{t}$ and another one from \textit{unexplored} area, i.e. $\mathbb{S} \backslash \mathcal{B}_{t}$. (Both selections are \textbf{random} in a given subset of food items). With above explanations, the image selection method we propose in this application is shown in Algorithm \ref{algo:select}.

\begin{algorithm}
\SetKwFunction{Select}{select}
\SetKwFunction{Percentile}{percentile}
\SetKwFunction{Kmeanspp}{k-means-pp}
\SetKwFunction{Random}{random}
\SetKwInOut{Input}{input}
\SetKwInOut{Output}{output}
\SetKwProg{Fn}{Function}{}{}

\Fn{\Select{$t, \mathcal{B}_{t}, \boldsymbol{p}^{t}$}}{
\Input{$t, \mathcal{B}_{t}, \boldsymbol{p}^{t}$}
\Output{$\mathcal{K}_{t}$}
\BlankLine
$\mathcal{K}_{t}=\emptyset$

\uIf{$t \leq 2$}{
$\mathcal{K}_{t} \leftarrow$ \Kmeanspp{$\mathbb{S}$, 10} \tcp{K-means++}
}
\Else{
\tcp{99th percentile (top $1\%$)}
threshold $\leftarrow$ \Percentile{$\boldsymbol{p}^{t}$, 99}\\
topSet $\leftarrow$ $\{s_{i} \in \mathbb{S}|p_{i}^{t} \geq $ threshold$\}$ \\
$\mathcal{K}_{t} \leftarrow$ [\Random{topSet}, \Random{$\mathbb{S}\backslash \mathcal{B}^{t}$}]
}
}
\caption{Images Selection Algorithm - \textit{select}}
\label{algo:select}
\end{algorithm}

\section{FoodDist: Food Image Embedding}
\label{sec:fooddist}

\begin{figure}
  \centering
  \includegraphics[width=0.5\columnwidth]{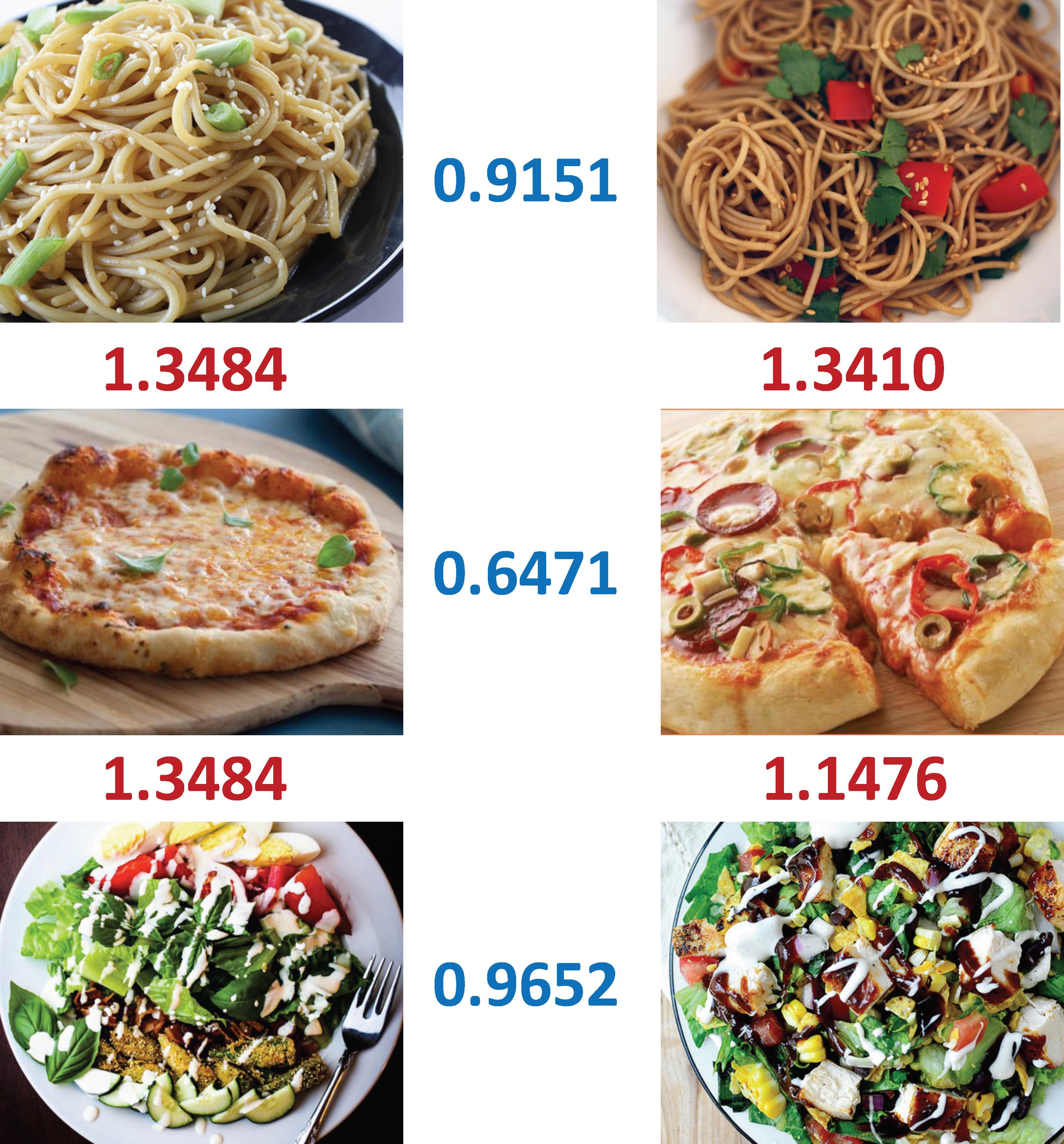}
  \caption{Euclidean embedding of FoodDist. This figure shows the pairwise euclidean distances between food images in the embedding. A distance of 0.0 means two food items are identical and a distance of 2.0 represents that the image contents are completely different. For this example, if the threshold is set to 1.0, then all the food images can be correctly classified.}
    \label{fig:distance}
\end{figure}

Formally, the goal of FoodDist is to learn a \textbf{feature extractor (embedding)} $f$ such that given an image $x$, $f(x)$ projects it to an $N$ dimensional feature vector for which the Euclidean distance to other such vectors will reflect the similarities between food images, as Fig.~\ref{fig:distance} shows. Formally speaking, if image $x_{1}$ is more similar to image $x_{2}$ than image $x_{3}$, then $\| f(x_{1}) - f(x_{2}) \| < \| f(x_{1}) - f(x_{3}) \|$.

We build FoodDist based on recent advances in deep Convolutional Neural Networks (CNN), which provide a powerful framework for automatic feature learning. Traditional feature representations for images are mostly hand-crafted, and were used with feature descriptors, such as SIFT (Scale Invariant Feature Transform)~\cite{lowe2004distinctive}, which aims for invariance to changes in object scale and illumination, thereby improving the generalizability of the trained model. However, in the face of highly diverse image characteristics, the one-size-fits-all feature extractor performs poorly. In contrast, deep learning adapts the features to particular image characteristics and extracts features that are most discriminative in the given task~\cite{razavian2014cnn}.

As we present below, a feature extractor for food images can be learned through classification and metric learning, or through multitask learning, which concurrently performs these two tasks. We demonstrate that the proposed multitask learning approach enjoys the benefits of both classification and metric learning, and achieves the best performance.

\subsection{Learning with classification}

One common way to learn a feature extractor for labeled data is to train a neural network that performs classification (i.e., mapping input to labels), and takes the output of a hidden layer as the feature representations; specifically, using a feedforward deep CNN with $n$-layers (as the upper half of the Fig.~\ref{fig:FoodDist} shows):

\begin{equation}
F(x) = g_n\left( g_{n-1}\left( \dots g_i( \dots g_1(x) \dots )\right)\right)
\end{equation}

where $g_i(.)$ represents the computation of $i$-th layer (e.g., convolution, pooling, fully-connected, etc.), and $F(x)$ is the output class label. The difference between the output class label and the ground truth (i.e., the error) is back-propagated throughout the whole network from layer $n$ to the layer $1$. We can take the output of the layer $n-1$ as the feature representation of $x$, which is equivalent to having a feature extractor $f$ as:

\begin{equation}
f(x) = g_{n-1}\left( \dots g_i( \dots g_1(x) \dots )\right)
\end{equation}

Usually, the last few layers will be fully-connected layers, and the last layer $g_n(.)$ is roughly equivalent to a linear classifier that is built on the features $f(x)$~\cite{Goodfellow-et-al-2016-Book}. Therefore, $f(x)$ is discriminative in separating instances under different categorical labels, and the Euclidean distances between normalized feature vectors can reflect the similarities between images.

\subsection{Metric Learning}
\begin{figure*}
  \centering
  \includegraphics[width=\columnwidth]{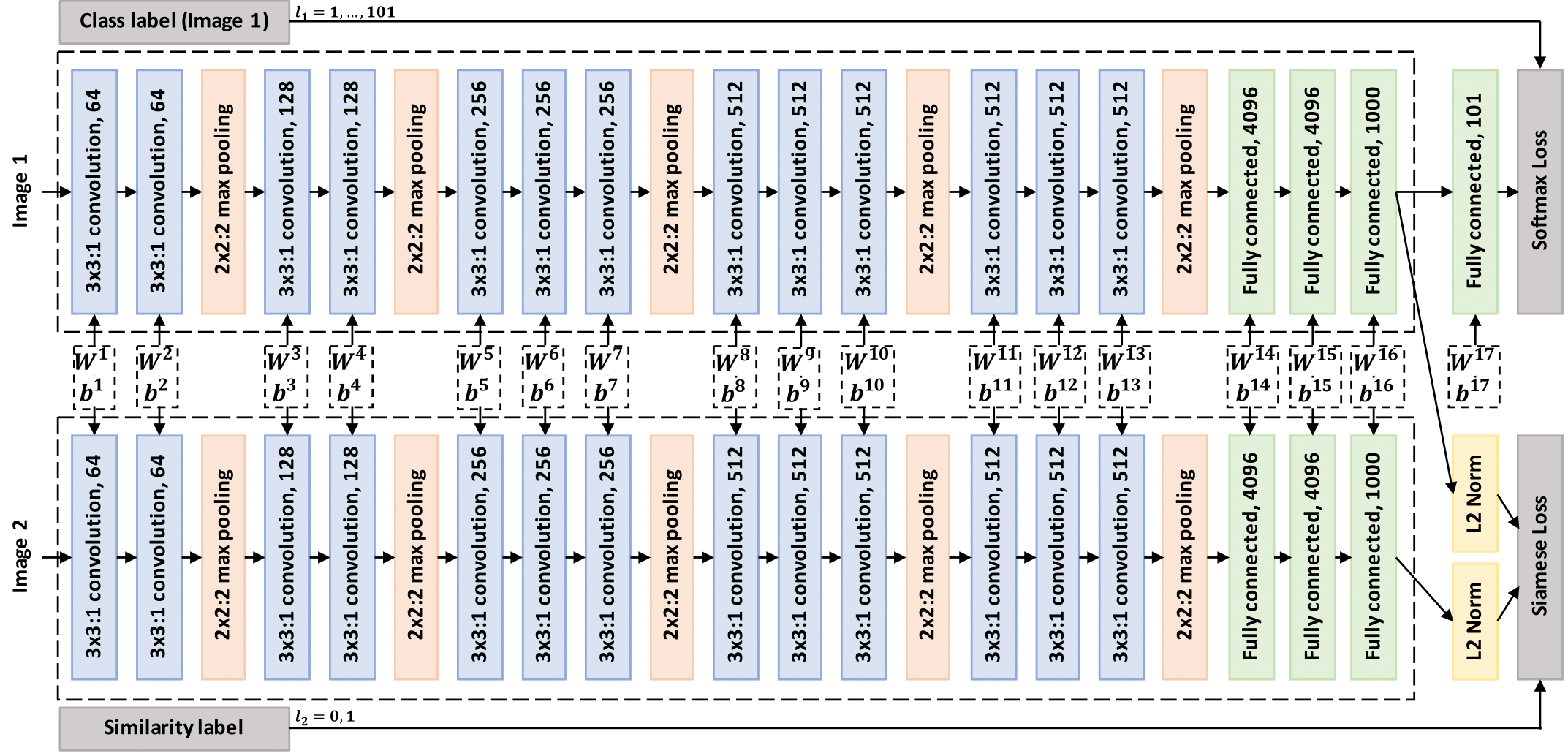}
  \caption{Multitask learning structure of FoodDist. Different types of layers are denoted by different colors. The format of each type of layer: Convolution layer: [\underline{receptive field size}:\underline{step size} ..., \underline{\#channels}]; Pooling layer: [\underline{pooling size}:\underline{step size} ...]; Fully connected layer: [..., \underline{output dimension}].}
    \label{fig:FoodDist}
    \vspace{-3mm}
\end{figure*}

Different from the classification approach, where the feature extractor is a by-product, metric learning proposes to learn the distance embedding directly from  the paired inputs of similar and dissimilar examples. Prior work~\cite{yang2015plateclick} used a \textit{Siamese network} to learn a feature extractor for food images. The structure of a Siamese network resembles that in Fig.~\ref{fig:FoodDist} but without \textit{Class label}, \textit{Fully connected, 101} and \textit{Softmax Loss} layers. The inputs to the Siamese network are pairs of food images $x_{1}, x_{2}$. The images pass through  CNNs with shared weights and the output of each network is regarded as the feature representation, i.e., $f(x_{1})$ and $f(x_{2})$, respectively. Our goal is for $f(x_{1})$ and $f(x_{2})$ to have a small distance value (close to 0) if $x_{1}$ and $x_{2}$ are similar food items;  otherwise, they should have a larger distance value. The value of contrastive loss is then back-propagated to optimize the Siamese network:

\begin{equation}
\mathcal{L}(x_{1}, x_{2}, l) = \frac{1}{2} l D^2 + \frac{1}{2} (1 - l) \max\left(0, m - D\right)^2
\end{equation}

where similarity label $l \in \{0, 1\}$ indicates whether the input pair of food items $x_{1}$, $x_{2}$ are similar or not ($l = 1$ for similar, $l = 0$ for dissimilar), $m > 0$ is the margin for dissimilar items and $D$ is the Euclidean distance between $f(x_{1})$ and $f(x_{2})$ in embedding space. Minimizing the contrastive loss will pull similar pairs together and push dissimilar pairs farther away (larger than a margin $m$) and it exactly matches the goal.

The major advantage of metric learning is that the network will be directly optimized for our final goal, i.e., a robust distance measure between images. However, as shown in the model benchmarks, using the pairwise information alone does not improve the embedding performance as the process of sampling pairs loses the label information, which is arguably more discriminative than (dis)similar pairs.

\subsection{Multitask Learning: concurrently optimize both tasks}

Both methods above have their pros and cons. Learning with classification leverages the label information, but the network is not directly optimized to our goal. As a result, although the feature vectors are learned to be separable in the linear space, the intra- and inter- categorical distances might still be unbalanced. On the other hand, metric learning is explicitly optimized for our final objective by pushing the distances between dissimilar food items apart beyond a margin $m$. Nevertheless, sampling the similar or dissimilar pairs loses valuable label information. For example, given a pair of items with different labels, we only consider the dissimilarity between the two categories they belong to, but overlook the fact that each item is also different from the remaining $n-2$ categories, where $n$ is the total number of categories.

In order to leverage the benefits of both tasks, we propose a multitask learning design~\cite{Goodfellow-et-al-2016-Book} for FoodDist. The idea of multitask learning is to share part of the model across tasks so as to improve the generalization ability of the learned model~\cite{Goodfellow-et-al-2016-Book}. In our case, as Fig.~\ref{fig:FoodDist} shows, we share the parameters between the classification network and Siamese network, and optimize them simultaneously. We use the base structure of the Siamese network and share the upper CNN with a classification network where the output of the CNN is fed into a cascade of a fully connected layer and a softmax loss layer. The final loss of the whole network is the weighted sum of the softmax loss $\mathcal{L}_{\text{softmax}}$ and contrastive loss $\mathcal{L}_{\text{contrastive}}$:

\begin{equation}
\mathcal{L} = \omega \mathcal{L}_{\text{softmax}} + (1-\omega)\mathcal{L}_{\text{contrastive}}
\end{equation}

Our benchmark results (Section \ref{sec:fooddist_benchmarking}) suggest that the feature extractor built with multitask learning achieves the best of both worlds: it achieves the best  performance for both classification and Euclidean distance-based retrieval tasks. 


\section{Evaluation}
\label{sec:evaluation}

We conduct user testing for online learning framework and end-to-end recommender system (\textbf{Yum-me}), as well as offline evaluation for food image embedding model (\textbf{FoodDist}). Our hypothesis for the evaluations are summarized below:

\begin{itemize}
    \item \textbf{H1}: Our online learning framework learns more accurate food preference profile than baseline approaches.
    \item \textbf{H2}: FoodDist generates better similarity measure for food images than state-of-the-art embedding models.
    \item \textbf{H3}: Yum-me makes more accurate nutritionally-appropriate meal recommendations than traditional survey as it integrates coarse-grained item filtering (provided by survey) with fine-grained food preference learned through adaptive elicitation.
\end{itemize}

In this section, we first present user testing results for online learning framework in Section \ref{sec:user_testing_online_learning}, then offline benchmark \textbf{FoodDist} model with a large-scale real-world food image dataset in Section \ref{sec:fooddist_benchmarking}, and finally discuss the results of end-to-end user testing in Section \ref{sec:yum-me user study}.

\subsection{User testing for online learning framework}
\label{sec:user_testing_online_learning}

In order to evaluate the accuracy of our online learning framework, we conducted a field study among 227 anonymous users recruited from social networks and university mailing lists. The experiment was approved by Institutional Review Board (ID: 1411005129) at Cornell University. All participants were required to use this system independently for three times. Each time the study consisted of following two phases:

\begin{itemize}
    \item \textit{Training Phase.} Users conducted the first $T$ iterations of food image comparisons, and the system learnt and elicited preference vector $\boldsymbol{p}^{T}$ based on the algorithms proposed in this paper or baseline methods, which will be discussed later. We randomly picked $T$ from set $\{5, 10, 15\}$ at the beginning but made sure that each user experienced different values of $T$ only once.
    \item \textit{Testing Phase.} After $T$ iterations of training, users entered the testing phase, which consisted of 10 rounds of pairwise comparisons. We picked testing images based on preference vector $\boldsymbol{p}^{T}$ that learnt from online interactions: One of them was selected from food area that user liked (i.e. item with top $1\%$ preference value) and the other one from the area that user disliked (i.e. item with bottom $1\%$ preference value) Both of the images were picked \textbf{randomly} among \textbf{unexplored} food items.
\end{itemize}

\subsubsection{Prediction accuracy}

In order to evaluate the effectiveness of \textit{user state update} and \textit{images selection} methods respectively, we conduct a 2-by-2 experiment in this section. For the \textit{user state update} method, we compare proposed \textit{Label propagation, Exponentiated Gradient} (\textbf{LE}) algorithm to \textit{Online Perceptron} (\textbf{OP}), and for the \textit{images selection} method, we compare proposed \textit{Exploration-Exploitation} (\textbf{EE}) algorithm to the \textit{Random Selection} (\textbf{RS}). Specifically, four frameworks presented below are evaluated. Users encountered them \textbf{randomly} when they logged into the system:

\textbf{LE+EE:} This is the online learning algorithm proposed in this paper that combines the ideas of \textbf{L}abel propagation, \textbf{E}xponentiated Gradient algorithm for user state update and \textbf{E}xploitation-\textbf{E}xploration strategy for images selection.

\textbf{LE+RS:} This algorithm retains our method for user state update (\textbf{LE}) but \textbf{R}andom \textbf{S}elect images to present to user without any exploitation or exploration.   

\textbf{OP+EE:} As each item is represented by \textit{1000 dim} feature vector, we can adopt the idea of regression to tackle this online learning problem (i.e. learning weight vector $\boldsymbol{w}$ such that $\boldsymbol{w}f(x_{i})$ is higher for item $i$ that user prefer). Hence, we compare our method with \textbf{O}nline \textbf{P}erceptron algorithm that updates $\boldsymbol{w}$ whenever it makes error, i.e. if $y_{i}\boldsymbol{w}f(x_{i}) \leq 0$, assign $\boldsymbol{w} \leftarrow \boldsymbol{w} + y_{i}\boldsymbol{w}f(x_{i})$, where $y_{i}$ is the label for item $i$ (pairwise comparison is regarded as binary classification such that the food item that user select is labeled as +1 and otherwise -1). In this algorithm, we retain our strategy of images selection (i.e. \textbf{EE}).

\textbf{OP+RS:} The last algorithm is the \textbf{O}nline \textbf{P}erceptron mentioned above but with \textbf{R}andom images \textbf{S}election strategy.

Among 227 participants in our study, 58 of them finally used algorithm \textbf{LE+EE}, 57 used \textbf{OP+RS}. For the rest of users (112), half of them (56) tested \textbf{OP+EE} and the other half (56) tested \textbf{LE+RS}. Overall, the participants for different algorithms are totally random so that the performances of different models are directly comparable.

\begin{figure} [t]
\centering
\includegraphics[width=0.6\columnwidth]{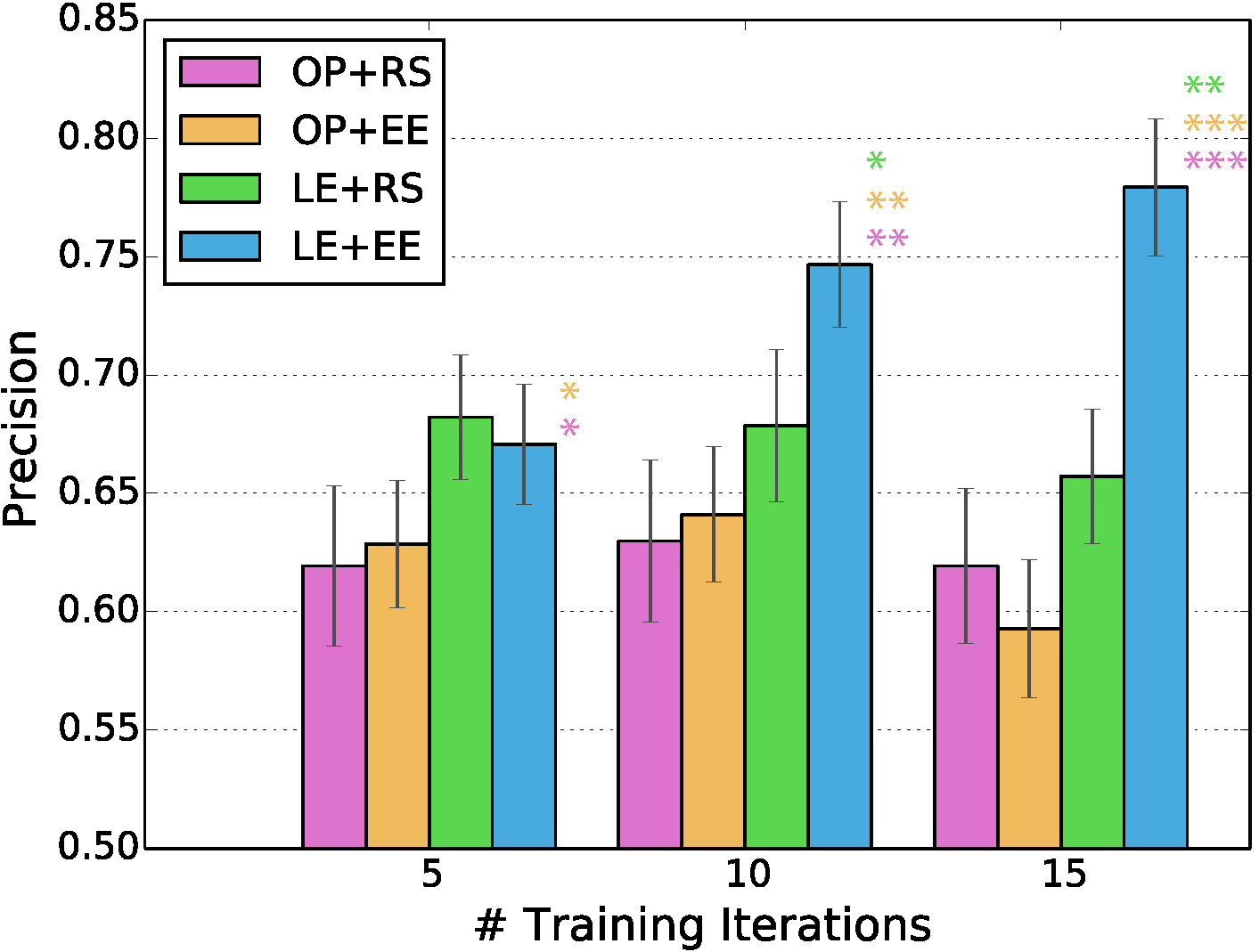}
\caption{Prediction accuracy for different algorithms in various training settings (asterisks represent different levels of statistical significance: $***:p < 0.001$, $**:p<0.01$, $*:p<0.05$).}
\label{fig:prediction}
\end{figure}

\begin{figure} [t]
\centering
\includegraphics[width=0.6\linewidth]{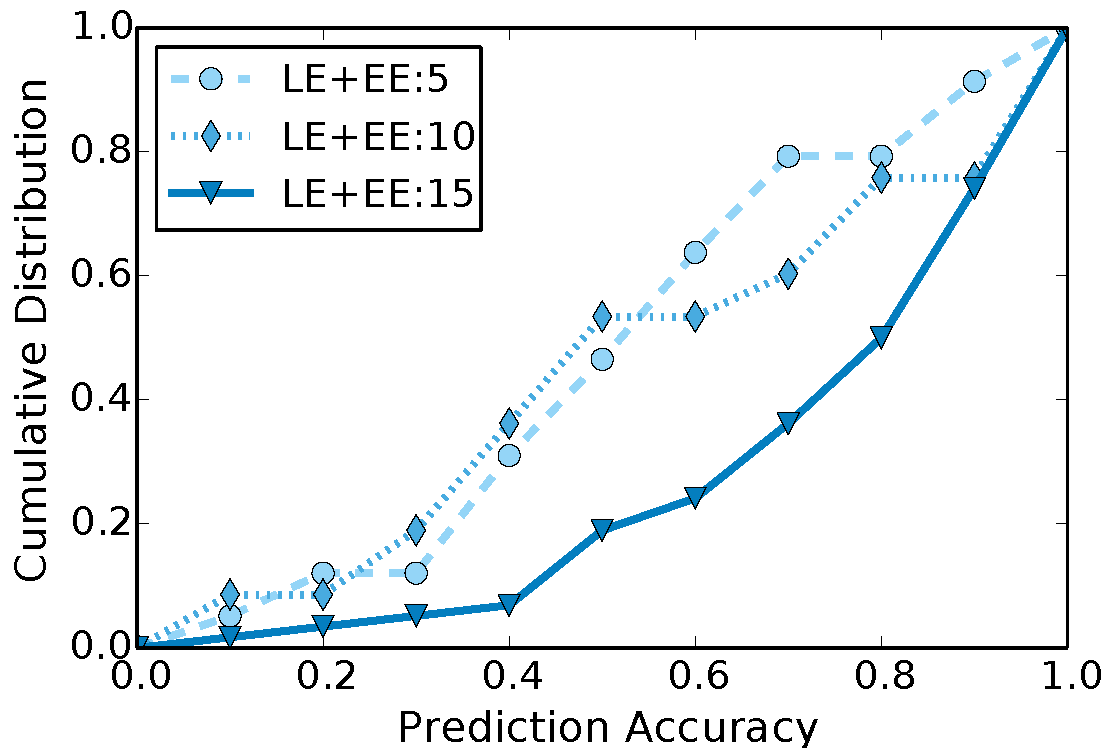}
\caption{Cumulative distribution of prediction accuracy for \textit{LE+EE} algorithm (Numbers in the legend represent the number of training iterations (i.e. values of $T$)).}
\label{fig:distribution}
\end{figure}

\begin{figure*}[t]
\centering
\subfigure[\# training iterations: 5]{
\includegraphics[width=.315\columnwidth]{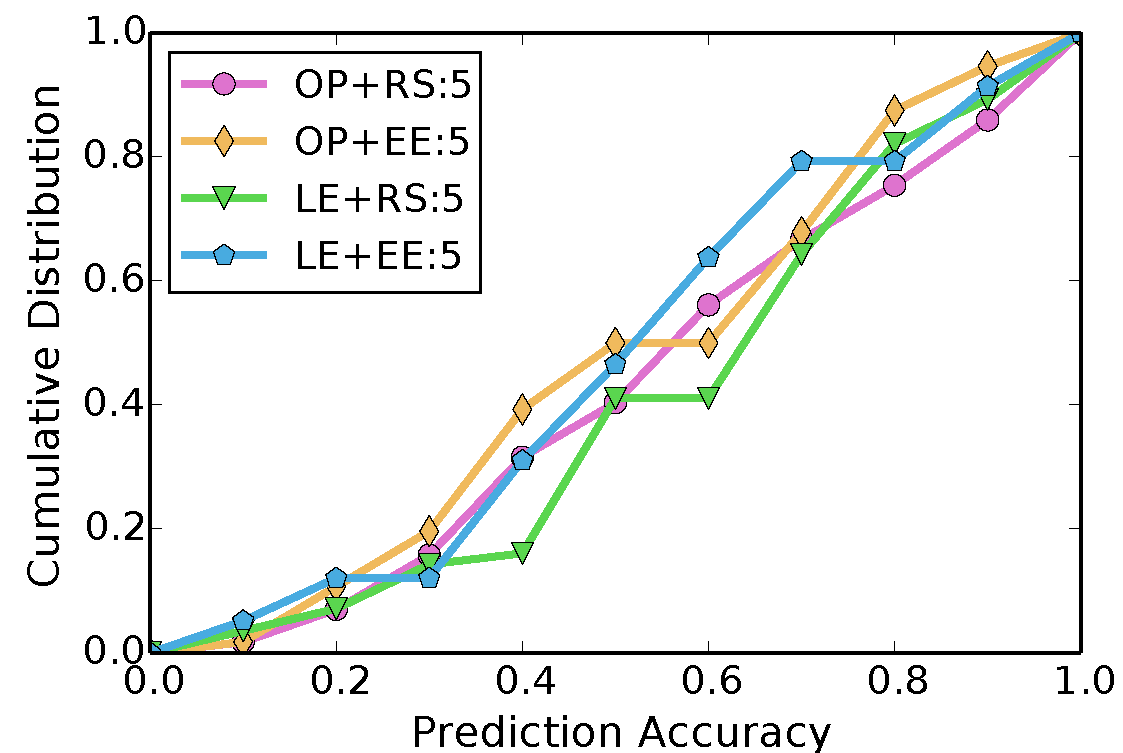}
   \label{Fig:distribution_2}
 }
\subfigure[\# training iterations: 10]{
\includegraphics[width=.315\columnwidth]{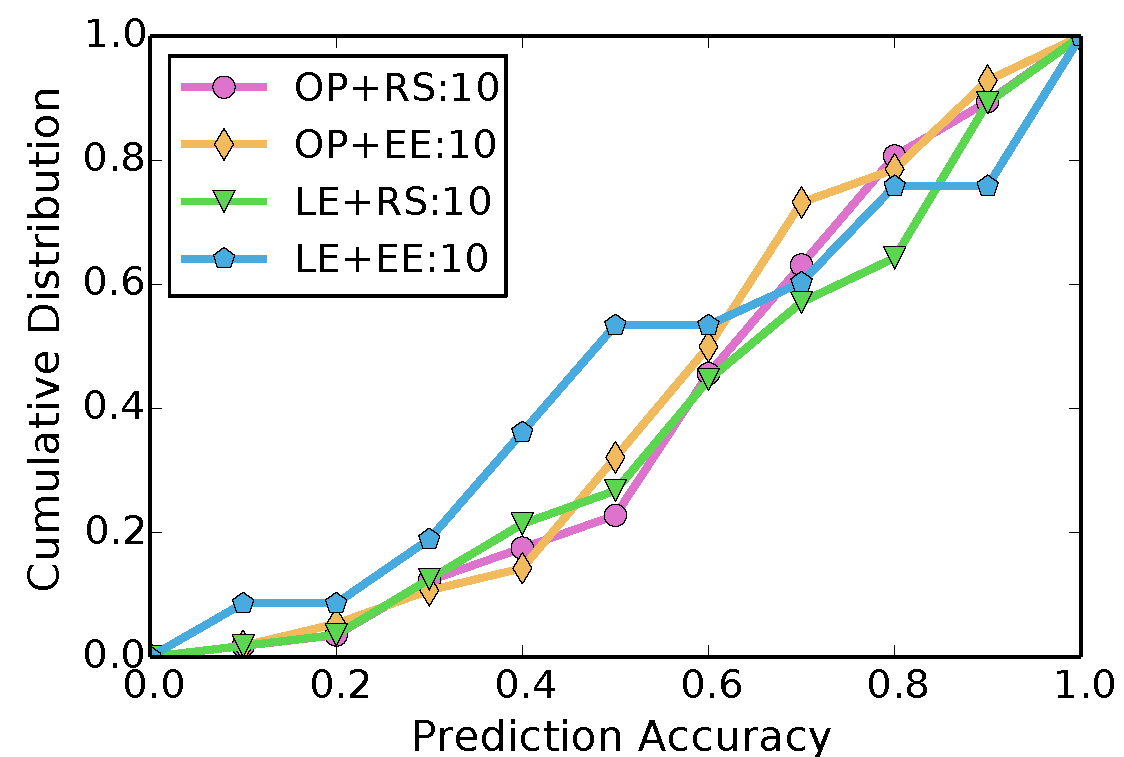}
   \label{Fig:distribution_3}
 }
\subfigure[\# training iterations: 15]{
\includegraphics[width=.315\columnwidth]{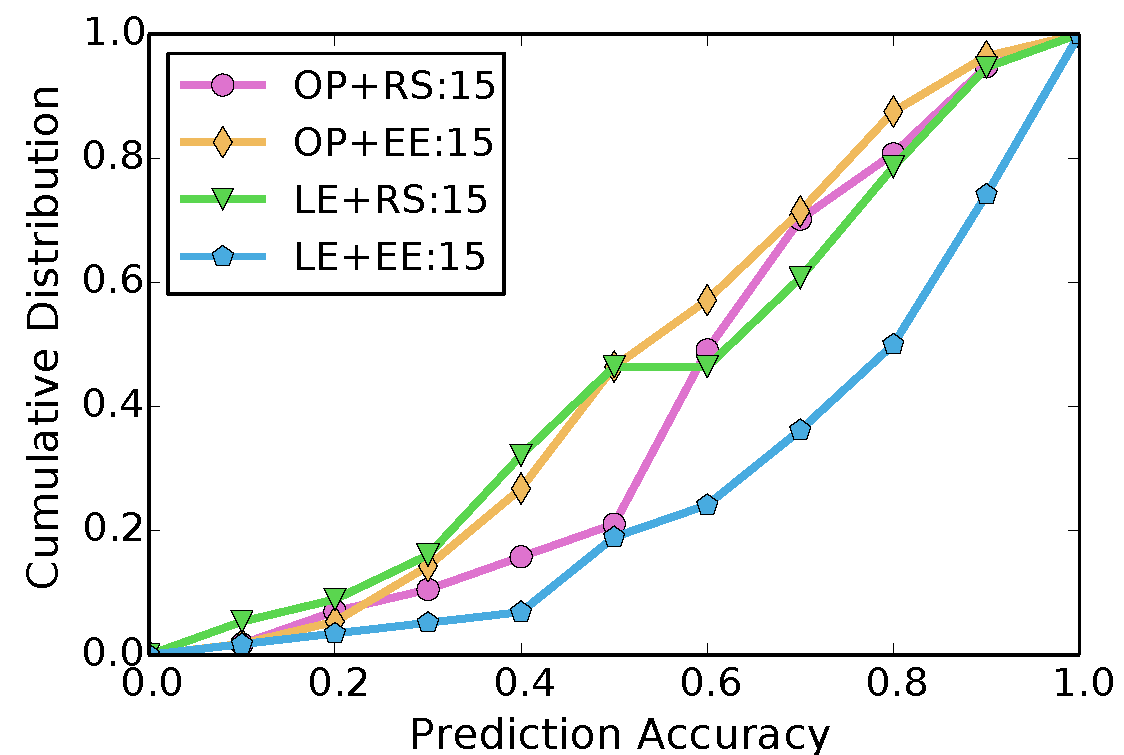}
   \label{Fig:distribution_4}
 }
\caption{Comparison of cumulative distribution of prediction accuracy across different algorithms.}
\label{fig:distribution_all}
\end{figure*}

After all users going through the training and testing phases, we calculate the prediction accuracy of each \textbf{individual user} and aggregate them based on the context that they encountered (i.e. the number of training iterations $T$ and the algorithm settings mentioned above). The prediction accuracies and their cumulative distributions are shown in Fig.~\ref{fig:prediction}, \ref{fig:distribution} and \ref{fig:distribution_all} respectively.

\textbf{Length effects of training iterations.} As shown in Fig.~\ref{fig:prediction} and Fig.~\ref{fig:distribution}, the prediction accuracies of our online learning algorithm are all significantly higher than the baselines.The algorithm performance is further improved with longer training period. As is clearly shown in Fig.~\ref{fig:distribution}, when the number of training iterations reaches 15, about half of the users will experience the prediction accuracy that exceeds $80\%$, which is fairly promising and decent considering small number of interactions that system elicited from scratch. The results above justify that the online preference learning algorithm can adjust itself to explore users' preference area as more information is available from their choices. For the task of item-based food preference bootstrapping, our system can efficiently balance the exploration-exploitation while providing reasonably accurate predictions.

\textbf{Comparisons across different algorithms.} As mentioned previously, we compared our algorithm with several obvious alternatives. As shown in Fig.~\ref{fig:prediction} and Fig.~\ref{fig:distribution_all}, none of these algorithms works very well and the accuracy of prediction is actually decreasing as the user provides more information. Additionally, as is shown in Fig.~\ref{fig:distribution_all}, our algorithm has particular advantages when users are making progress (i.e. the number of training iterations reaches 15). The reason why these techniques are not suited for our application is mainly due to the following limitations: 

\textit{Random Selection.} Within a limited number of interactions, random selection can not maintain the knowledge that it has already learned about the user (exploitation), nor explore unknown areas (exploration). In addition, it's more likely that the system will choose food items that are very similar to each other and thus hard for the user to make decisions. Therefore, after short periods of interactions, the system is messed up, and the performance degrades.

\textit{Underfitting.} The algorithm that will possibly have the underfitting problem is the online perceptron (\textbf{OP}). For our application, each food item is represented by \textit{1000 dim} feature vector and \textbf{OP} is trying to learn a separate hyperplane based on a limited number of training data. As each single feature is directly derived from deep neural network, the linearity assumptions made by perceptron might yield wrong predictions for the dishes that haven't been explored before.

\subsubsection{System efficiency}

\begin{figure*}[t]
\centering
\subfigure[User Response Time]{
\includegraphics[width=.97\columnwidth]{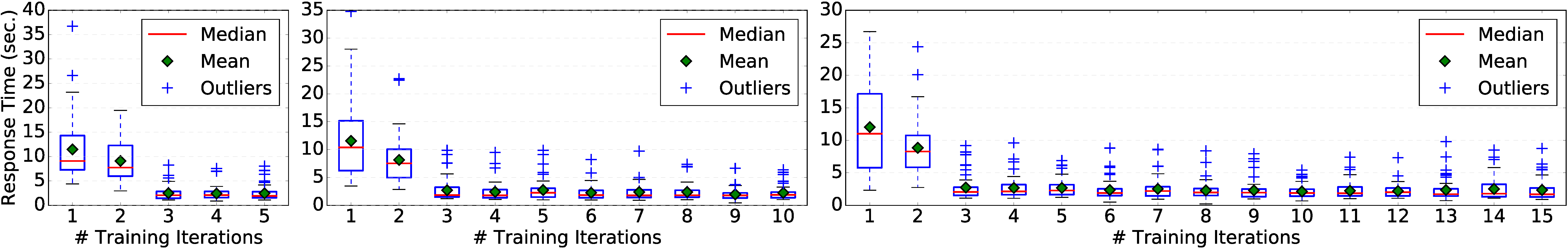}
   \label{fig:response_time}
 }
\subfigure[System Execution Time]{
\includegraphics[width=.98\columnwidth]{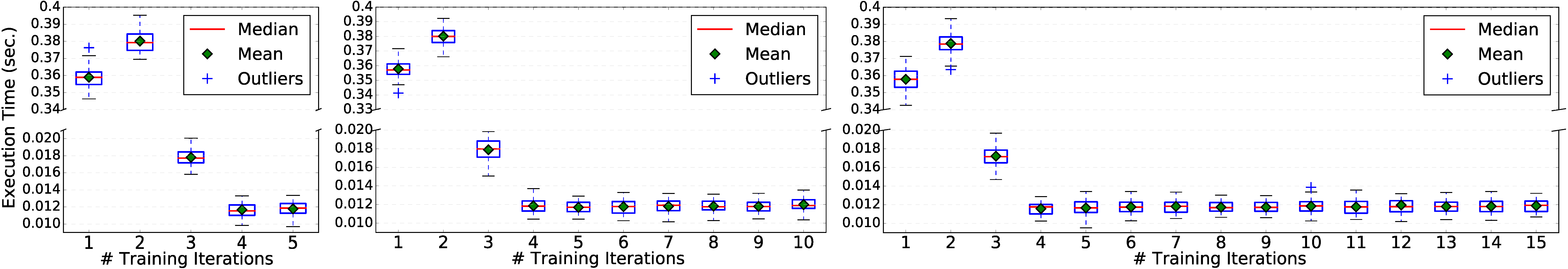}
   \label{fig:execution_time}
 }
\caption{Timestamp records for user response time and system execution time.}
\label{fig:time}
\end{figure*}

As another two aspects of online preference elicitation system, computing efficiency and user experience are also very important metrics for system evaluation. Therefore, we recorded the program execution time and user response time as a lens into the real-time performance of the online learning algorithm. As shown in Fig.~\ref{fig:execution_time}, the program execution time is about $0.35s$ for the first two iterations and less than $0.025s$ for the iterations afterwards\footnote{Our web system implementation is based on Amazon EC2 t2-micro Linux 64-bit instance}. Also, according to Fig.~\ref{fig:response_time}, the majority of users can make their decisions in less than $15s$ for the task of comparison among ten food images while the payload for the pairwise comparison is less than $2-3s$. As a final cumulative metric for the system overhead, it is shown in Table.~\ref{tbl:time} that even for $15$ iterations of training, users can typically complete the whole process within 53 seconds, which further justify that our online learning framework is light-weight and user-friendly in efficiently eliciting food preference.

\begin{table} [h]
\caption{Average time to complete training phase.}
\label{tbl:time}

\begin{tabular}{|c|c|c|}
\hline
\# Iter: 5 & \# Iter: 10 & \# Iter: 15 \\ \hline
28.75s                            & 39.74s                             & 53.22s                             \\ \hline
\end{tabular}

\end{table}

\subsubsection{User qualitative feedback} After the study, some participants send us emails regarding their experiences towards the adaptive visual interface. Most of the comments reflect the participants' satisfactions and that our system is able to engage the user throughout the elicitation process. For example, ``\textit{Now I'm really hungry and want a grilled cheese sandwhich!}'', ``\textit{That was fun seeing tasty food at top of the morning.}'' and ``\textit{Pretty cool tool.}''. However, they also highlight some limitations of our current prototype. For example, ``\textit{I am addicted to spicy food and it totally missed it. There may just not be enough spicy alternatives in the different dishes to pick up on it.}'' points out that the prototype is limited in the size of the food database.

\subsection{Offline benchmarking for FoodDist}
\label{sec:fooddist_benchmarking}

We develop FoodDist and baseline models (Section \ref{sec:fooddist}) using Food-101 training dataset, which contains 75,750 food images from 101 food categories (750 instances for each category)~\cite{bossard2014food}. To the best of our knowledge, Food-101 is the largest and most challenging publicly available dataset for food images. We implement models using Caffe~\cite{jia2014caffe} and experiment with two CNN architectures in our framework: AlexNet~\cite{krizhevsky2012imagenet}, which won the first place at ILSVRC2012 challenge, and VGG~\cite{Simonyan14c}, which is the state-of-the-art CNN model. The inputs to the networks are image crops of sizes $224 \times 224$ (VGG) or $227 \times 227$ (AlexNet). They are randomly sampled from a pixelwise mean-subtracted image or its horizontal flip. In our benchmark, we train four different feature extractors: AlexNet+Learning with classification (\textbf{AlexNet+CL}), AlextNet+Multitask learning (\textbf{AlexNet+MT}), VGG+Learning with classification (\textbf{VGG+CL}) and VGG+Multitask learning (\textbf{VGG+ML, FoodDist}). For the multitask learning framework, we sample the similar and dissimilar image pairs with 1:10 ratio from the Food-101 dataset based on the categorical labels to be consistent with the previous work~\cite{yang2015plateclick}. The models are fine-tuned based on the networks pre-trained with the ImageNet data. We
use Stochastic Gradient Decent with a mini-batch size of 64, and each network is trained for $10 \times 10^{4}$ iterations. The initial learning rate is set to 0.001 and we use a weight decay of 0.0005 and momentum of 0.9. 

We compare the performance of four feature extractors, including FoodDist, with the state-of-the-art food image analysis models using Food-101 testing dataset, which contains 25,250 food images from 101 food categories (250 instances for each category). The performance for \textbf{classification} and \textbf{retrieval} tasks are evaluated as follow:

\begin{itemize}
    \item \textbf{Classification:} We test the performance of using learned image features for classification. For the classification deep neural network in each of the models above, we adopt the standard 10-crop testing. i.e. the network makes a prediction by extracting ten patches (the four corner patches and the center patch in the original images and their horizontal reflections), and averaging the predictions at the softmax layer. The metrics used in this paper are Top-1 accuracy and Top-5 accuracy.
    \item \textbf{Retrieval:} We use a retrieval task to evaluate the quality of the euclidean distances between extracted features. Ideally, the distances should be smaller for similar image pairs and larger for dissimilar pairs. Therefore, as suggested by previous work~\cite{yang2015plateclick,yang2015beyond}, We check the nearest $k$-neighbors of each test image, for $k = 1, 2, ..., N$, where $N=25250$ is the size of the testing dataset, and calculate the Precision and Recall values for each $k$. We use mean Average Precision (mAP) as the evaluation metric to compare the performance. For every method, the Precision/Recall values are averaged over all the images in the testing set.
\end{itemize}

The classification and retrieval performance of all models are summarized in Table.~\ref{tab:acc} and Table.~\ref{tab:map} respectively. FoodDist performs the best among four models and is significantly better than the state-of-the-art approaches in both tasks. For the classification task, the classifier built on FoodDist features achieves 83.09\% Top-1 accuracy, which significantly outperforms the original RFDC~\cite{bossard2014food} model and the proprietary GoogLeNet model~\cite{meyers2015im2calories}; For the retrieval task, FoodDist doubles the mAP value reported by previous work~\cite{yang2015plateclick} that only used the AlexNet and siamese network architecture. The benchmark results demonstrate that FoodDist features possess high generalization ability and the euclidean distances between feature vectors reflect the similarities between food images with great fidelity. In addition, as we can observe from both tables, the multitask learning based approach always performs better than learning with classification for both tasks no matter which CNN is used. This further justifies the proposed multitask learning approach and its advantage of incorporating both label and pairwise distance information that makes the learned features more generalizable and meaningful in the euclidean distance embedding.

\begin{table}[h]
\caption{Model performance of classification task. $*$ represents state-of-the-art approach and bold text indicates the method with the best performance.}
\label{tab:acc}
\begin{tabular}{|c | c | c |} 

 \hline
 Method & Top-1 ACC (\%) & Top-5 ACC(\%) \\
 \hline\hline
 RFDC$*$~\cite{bossard2014food} & 50.76\% & $--$ \\ 
 \hline
 GoogleLeNet$*$~\cite{meyers2015im2calories} & 79\% & $--$ \\
 \hline\hline
 AlexNet+CL & 67.63\% & 89.02\% \\
 \hline
 AlexNet+MT & 70.50\% & 90.36\% \\
 \hline
 VGG+CL & 82.48\% & 95.70\% \\
 \hline
 \makecell{VGG+MT \textbf{(FoodDist)}} & \textbf{83.09\%} & \textbf{95.82\%} \\
 \hline

\end{tabular}
\end{table}

\begin{table}[h]
\caption{Model performance of retrieval task. $*$ represents state-of-the-art approach and bold text indicates the method with the best performance. (Note: The mAP value that we report for Food-CNN is higher because we use pixel-wise mean subtraction while the original paper only used per-channel mean subtraction.)}
\label{tab:map}
 \begin{tabular}{|c | c |} 
 
 \hline
 Method & mean Average Precision (mAP) \\
 \hline\hline
 Food-CNN$*$~\cite{yang2015plateclick} & 0.3084 \\ 
 \hline\hline
 AlexNet+CL & 0.3751 \\
 \hline
 AlexNet+MT & 0.4063 \\
 \hline
 VGG+CL & 0.6417 \\
 \hline
 \makecell{VGG+MT \textbf{(FoodDist)}} & \textbf{0.6670} \\
 \hline
\end{tabular}
    
\end{table}

\subsection{End-to-end user testing}
\label{sec:yum-me user study}

\begin{figure*}
  \centering
  \includegraphics[width=\columnwidth]{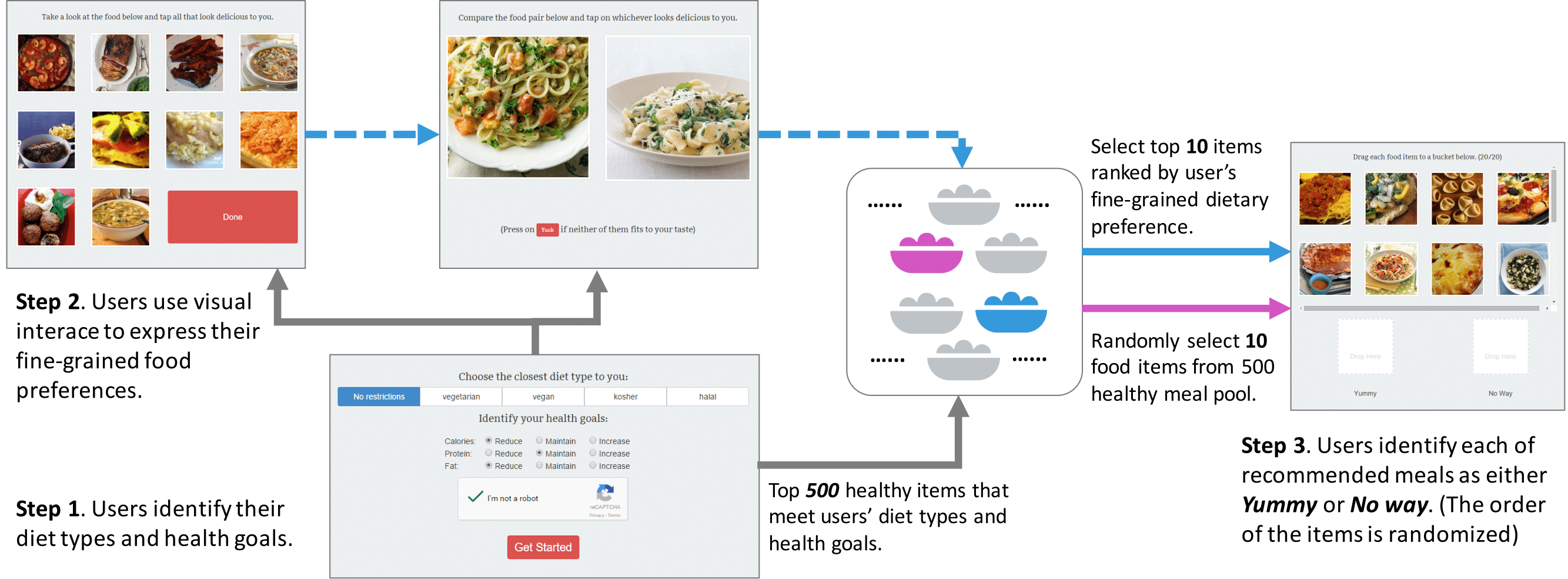}
  \caption{User study workflow for personalized nutrient-based meals recommendation system. We compare Yum-me (blue arrows) with the baseline method (violet arrow) that makes recommendations solely based on nutritional facts and dietary restrictions.}
    \label{fig:userstudy}
\end{figure*}

\begin{figure}
  \centering
  \includegraphics[width=0.6\columnwidth]{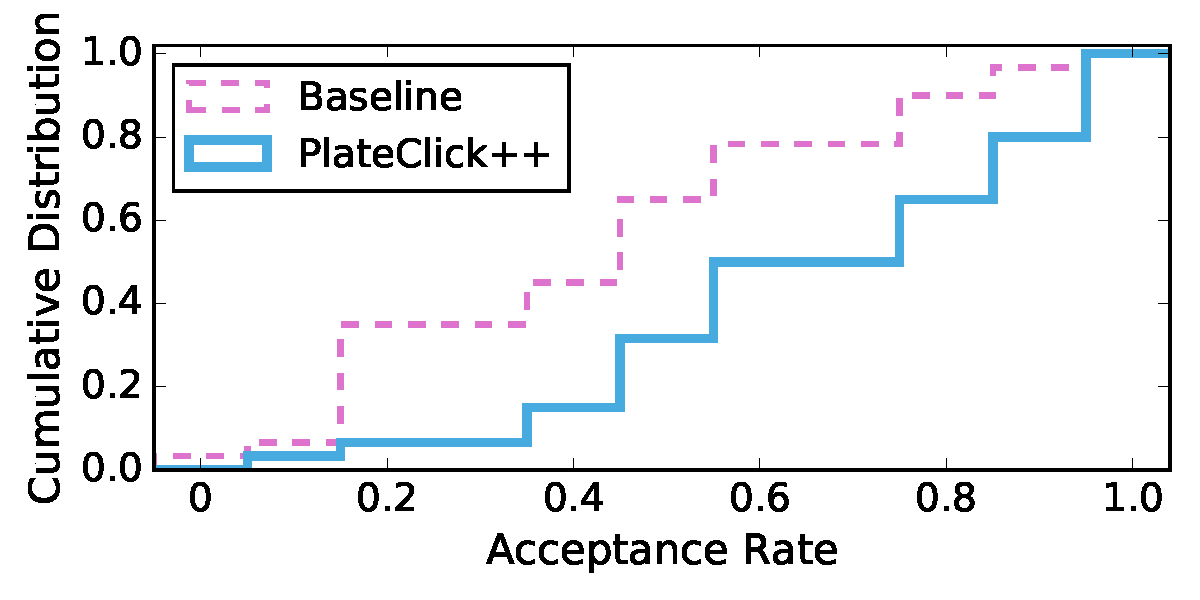}
  \caption{Cumulative distribution of acceptance rate for both recommender systems}
    \label{fig:acceptance}
\end{figure}

\begin{figure}
  \centering
  \includegraphics[width=0.7\columnwidth]{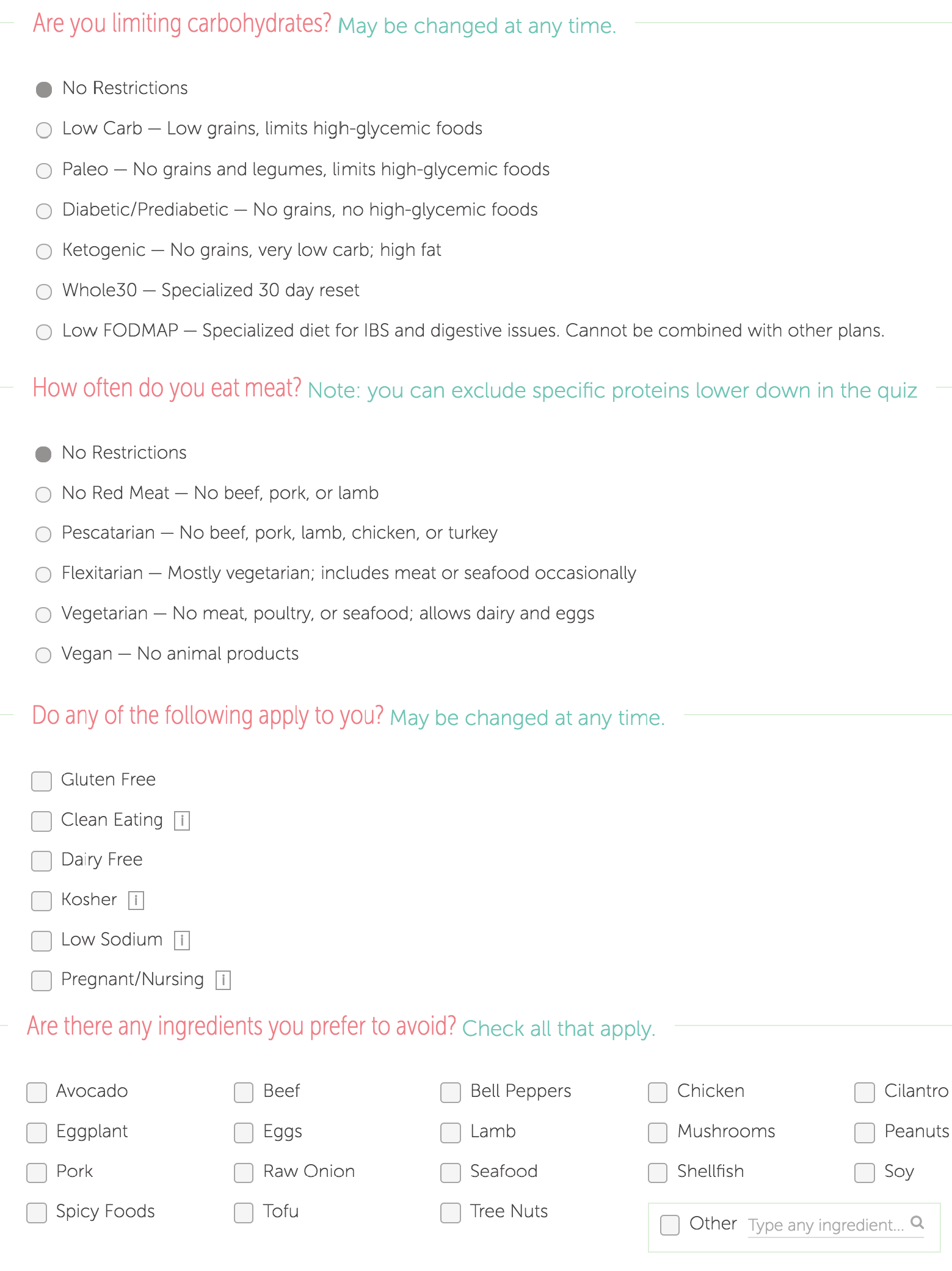}
  \caption{The survey used for user onboarding of PlateJoy. The questions are up-to-date at the time of paper writing, and we only include top four questions for illustration purpose.}
    \label{fig:survey}
\end{figure}

We conducted end-to-end user testing to validate the efficacy of Yum-me recommendations. We recruited 60 participants through the university mailing list, Facebook, and Twitter. The goal of the user testing was to compare Yum-me recommendations with a widely-used user onboarding approach, i.e. a traditional food preference survey (A sample survey used by PlateJoy is shown in Fig.~\ref{fig:survey}). As Yum-me is designed for scenarios where no rating or food consumption history is available (which is common when the user is new to a platform or is visiting nutritionist's office), collaborative filtering algorithm that has been adopted by many state-of-the-art recommenders is not directly comparable to our system.

In this study, we used a within-subjects study design in which each participant expressed their opinions regarding the meals recommended by both of the recommenders, and the effectiveness of the systems were compared on a per-user basis.

\subsubsection{Study Design}
 
We created a traditional recommendation system by randomly picking $N$ out of $M$ meals in the candidate pool to recommend to the users. The values of $N$ and $M$ are controlled such that $N=10, M=500$ for both Yum-me and the traditional baseline.  The user study consists of three phases, as Fig.~\ref{fig:userstudy} shows: (1) Each participant was asked to indicate their diet type and health goals through our basic user survey. (2) Each participant was then asked to use the visual interface. (3) 20 meal recommendations were arranged in a random order and presented to the participant at the same time, where 10 of them are made by Yum-me, and the other 10 are generated by  the baseline. The participant was asked to express their opinion by dragging each of the 20 meals into either the \textbf{Yummy} or the \textbf{No way} bucket. To overcome the fact that humans would tend to balance the buckets if their previous choices were shown, the food item disappeared  after the user dragged it into a bucket. In this way, users were  not reminded of how many meals they had put into each bucket.

The user study systems were implemented as web services and participants accessed the study from desktop or mobile browsers. We chose a web service for its wide accessibility to the population, but we could easily fit Yum-me into other ubiquitous devices, as mentioned earlier.

\subsubsection{Participants}

The most common dietary choice among our 60 participants was \textit{No restrictions} (48), followed by \textit{Vegetarian} (9), \textit{Halal} (2) and \textit{Kosher} (1). No participants chose \textit{Vegan}. Participant preferences in terms of nutrients are summarized in Table.~\ref{tab:health}. For \textit{Calories} and \textit{Fat}, the top two goals were \textit{Reduce} and \textit{Maintain}. For \textit{Protein}, participants tended to choose either \textit{Increase} or \textit{Maintain}. 
For health goals, the top four participant choices were \textit{Maintain calories-Maintain protein-Maintain fat} (20), \textit{Reduce calories-Maintain protein-Reduce fat} (10), \textit{Reduce calories-Maintain protein-Maintain fat} (10) and \textit{Reduce calories-Increase protein-Reduce fat} (5). The statistics match well with the common health goals among the general population, i.e. people who plan to control weight and improve sports performance tend to reduce the intake calories and fat, and increase the amount of protein. 

\begin{table}[h]
\caption{Statistics of health goals among 60 participants. Unit: number of participants.}
\label{tab:health}

 \begin{tabular}{| c |c | c | c |} 
 
 \hline
 Nutrient & Reduce & Maintain & Increase \\
 \hline\hline
 Calories & 30 & 28 & 2\\ 
 \hline
 Protein & 1 & 44 & 15 \\
 \hline
 Fat & 23 & 36 & 1 \\
 \hline
\end{tabular}
    
\end{table}
\vspace{-3mm}

\subsubsection{Quantitative analysis}

We use a quantitive approach to demonstrate that: (1) Yum-me recommendations yield higher meal acceptance rates than traditional approaches; and  (2) Meals recommended by Yum-me satisfy users' nutritional needs. 

In order to show higher meal acceptance rates, we calculated the participant acceptance rate of meal recommendations as: 

\[\frac{\text{\# Meals in Yummy bucket}}{\text{\# Recommended meals}}.\]

The cumulative distribution of the acceptance rate is shown in Fig.~\ref{fig:acceptance}, and the average acceptance rate, Mean Absolute Error (MAE) and Root Mean Square Error (RMSE) of each approach are presented in Table.~\ref{tab:acceptance}. The results demonstrate that Yum-me significantly improves the quality of the presented food items. The per-user acceptance rate difference between two approaches was normally distributed\footnote{A Shapiro Wilk W test was not significant ($p=0.12$), which justifies that the difference is normally distributed.}, and a paired Student's t-test indicated a significant difference between the two methods ($p<0.0001$).\footnote{We also performed a non-parametric Wilcoxon signed-rank test and found a comparable result.}
 
\begin{figure}
  \centering
  \includegraphics[width=0.6\columnwidth]{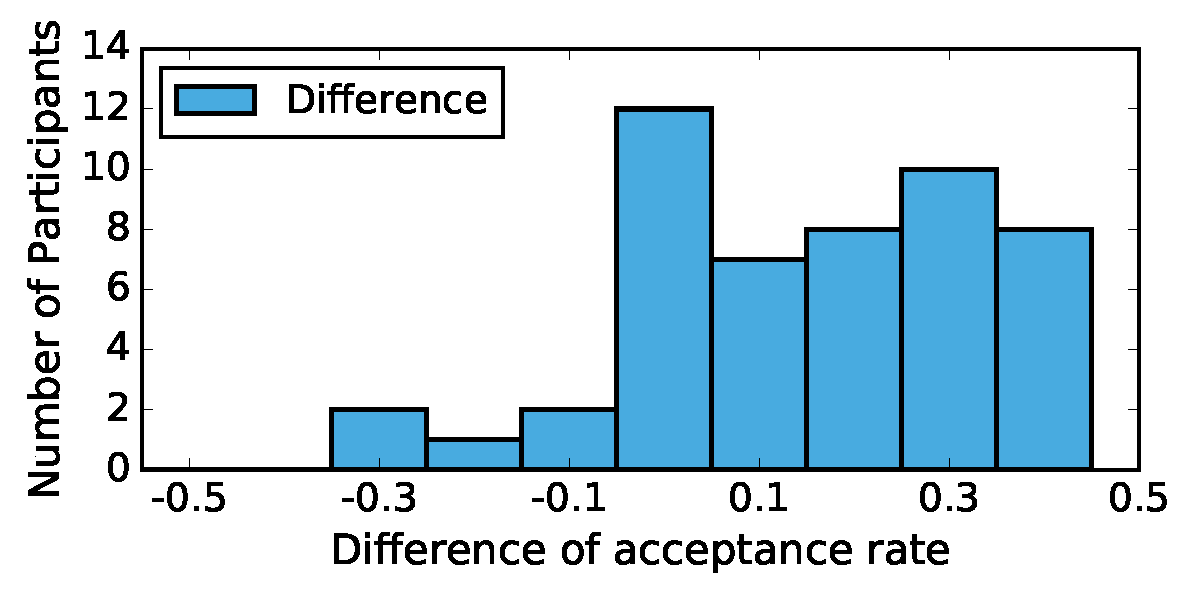}
  \caption{Distribution of the acceptance rate differences between two recommender systems.}
    \label{fig:difference}
\end{figure}

 To quantify the improvement provided by \textbf{Yum-me}, we calculated the difference between the acceptance rates of the two systems, i.e. \textit{$\text{difference}=\text{Yum-me acceptance rate}-\text{baseline acceptance rate}$}. The distribution and average values of the differences are presented in Fig.~\ref{fig:difference} and Table.~\ref{tab:acceptance} respectively. It is noteworthy that Yum-me outperformed the baseline by $42.63\%$ in terms of the number of preferred recommendations, which demonstrates its utility over the traditional meal recommendation approach. However, another observed phenomenon in Fig.~\ref{fig:difference} is that there are 12 users (20\%) with zero acceptance rate differences, which may due to the following two reasons: (1) Yum-me is not effective to this set of users, and it doesn't improve their preferences towards recommended food items. (2) As we didn't conduct participant control and filtering, some participants may not be well-involved in the study and randomly select or drag items.

\begin{table}[h]
\caption{Average Acceptance Rates (Avg. Acc.), Mean Absolute Error (MAE) and Root Mean Square Error (RMSE) between two systems. Paired t-test p-value (Avg. Acc.): $8.58 \times 10^{-10}$;}
\label{tab:acceptance}

 \begin{tabular}{| c | c | c |} 
 
 \hline
 Metric & Mean & SEM \\
 \hline\hline
 Yum-me Avg. Acc.  & \textbf{0.7250} & \textbf{0.0299}\\ 
 \hline
 Baseline Avg. Acc. & 0.5083 & 0.0341 \\
 \hline\hline
 Yum-me MAE  & \textbf{0.2750} & \textbf{0.0299}\\ 
 \hline
 Baseline MAE & 0.4916 & 0.0341 \\
 \hline\hline
 Yum-me RMSE  & \textbf{0.4481} & \textbf{0.0355}\\ 
 \hline
 Baseline RMSE & 0.6649 & 0.0290 \\
 \hline
\end{tabular}
    
\end{table}

\begin{figure}
	\centering
	\includegraphics[width=1.0\columnwidth]{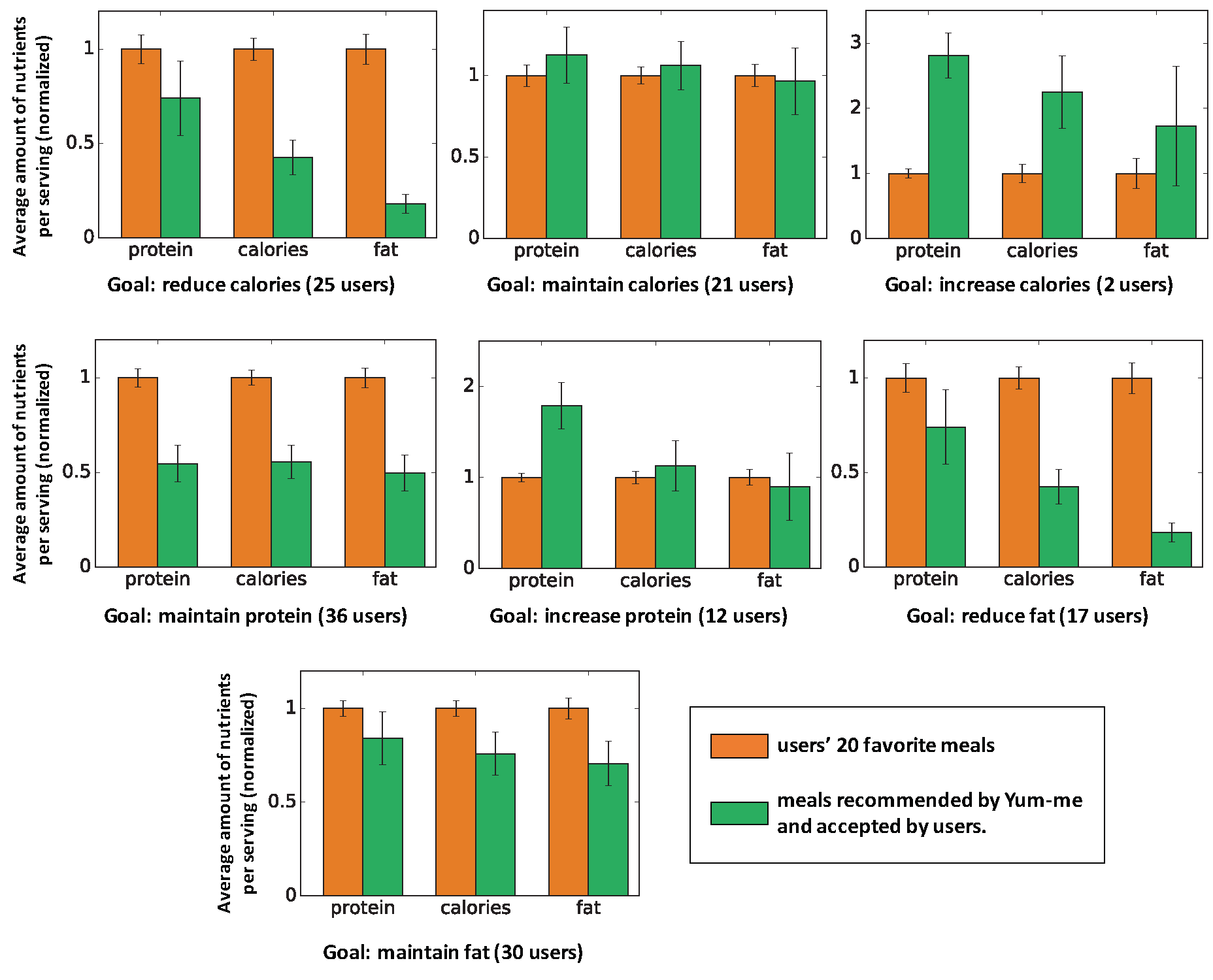}
	\caption{Nutritional facts comparison between paticipants' favorite meals and recommended (Yum-me) and accepted meals. The meal is accepted if it is dragged into the \textit{yummy} bucket. The mean values are normalized by the average amount of corresponding nutrient in the favorite meals (orange bar). (Only 7 out of 9 nutritional goals are used by at least one partipant). }
	\label{fig:nutrient}
\end{figure}

To examine meal nutrition, we compare the nutritional facts of paticipants' favorite meals with those of meals recommended (by Yum-me) and accepted (items dragged into the \textit{yummy} bucket) by the user. As shown in Fig.~\ref{fig:nutrient}, for users with same nutritional needs and no dietary restrictions, we calculate the average amount of protein, calories and fat (per-serving) in (1) their favorite 20 meals (as determined by our online learning algorithm), and (2) their recommended and accepted meals, respectively. The mean values presented in Fig.~\ref{fig:nutrient} are normalized by the average amount of corresponding nutrient in their favorite meals. The results demonstrate that using a relatively simple nutritional ranking approach, Yum-me is able to satisfy most of the nutritional needs set by the users, including \textit{reduce, maintain and increase calories}, \textit{increase protein}, and \textit{reduce fat}.  However, our system fails to meet two nutritional requirments, i.e. \textit{maintain protein} and \textit{maintain fat}. Our results also show where Yum-me recommendations result in unintended nutritional composition. For example, the goal of \textit{reducing fat} results in the reduction of \textit{protein} and \textit{calories}, and the goal of \textit{increasing calories} ends up increasing the \textit{protein} in meals. This is partially due to the inherent inter-dependance between nutrients and we leave further investigation of this issue to future work.

\begin{figure}
  \centering
  \includegraphics[width=0.7\columnwidth]{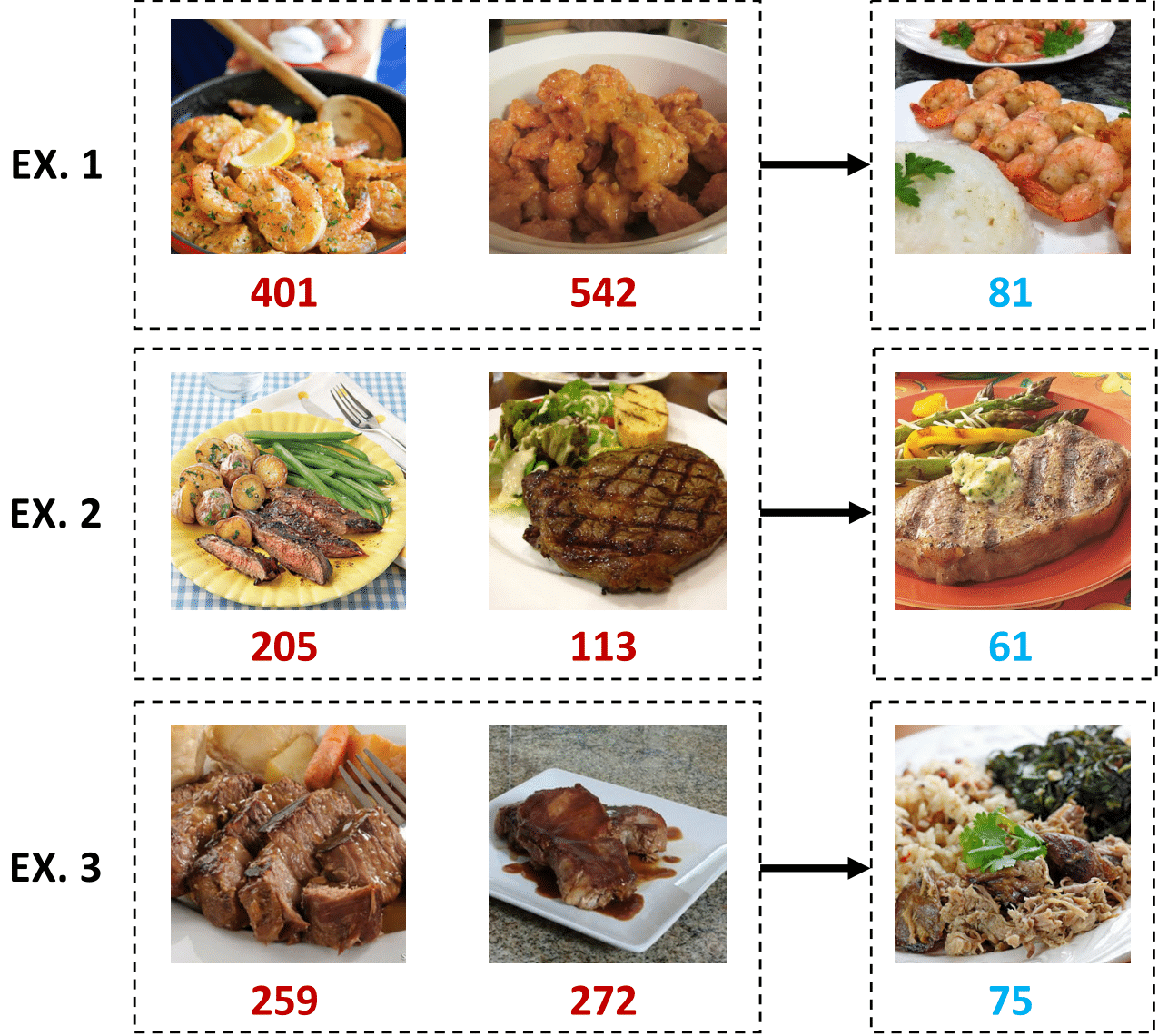}
  \caption{Qualitative analysis of personalized healthy meal recommendations. Images on the left half are sampled from users' top-20 favorite meals learned from Yum-me; Images on the right half are the meals presented to the user. The number under each food image represents the amount of calories for the dish, unit: \textit{kcal/serving}.}
    \label{fig:recsamples}
\end{figure}

\subsubsection{Qualitative analysis}


To qualitatively understand the personalization mechanism of Yum-me, we randomly pick 3 participants with no dietary restrictions and with the health goal of reducing calories. For each user, we select top-20 general food items the user likes most (inferred by the online learning algorithm). These food items played important roles in selecting the healthy meals to recommend to the user. To visualize this relationship,  among these top-20 items, we further select two food items that are most similar to the healthy items Yum-me recommended to the users and present three such examples in Fig.~\ref{fig:recsamples}. Intuitively, our system is able to recommend healthy food items that are \textit{visually similar} to the food items a user like, but the recommended items are of lower calories due to the use of healthier ingredients or different cooking styles. These examples showcase how Yum-me can project users' general food preferences to the domain of the healthy options, and find the ones that can most appeal to users.

\subsubsection{Error analysis}

\begin{figure}
  \centering
  \includegraphics[width=0.6\columnwidth]{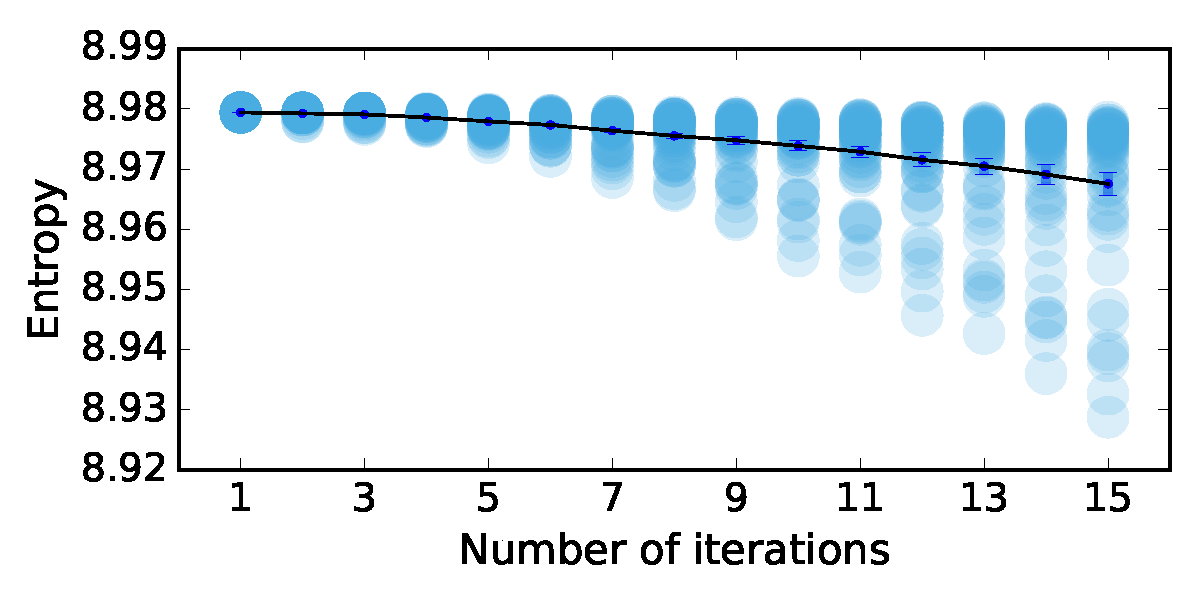}
  \caption{Entropy of preference distributions in different iterations of online learning. (Data from 48 users with no dietary restrictions)}
    \label{fig:entropy}
\end{figure}

Through a closer examination of the cases where our system performed, or did not perform, well, we observed a negative correlation between the entropy of the learned preference distribution $\mathbf{p}$ \footnote{Entropy of preference distribution: $H(\mathbf{p})=-\sum_{i}p_{i}\log p_{i}$} and the improvement of Yum-me over the baseline ($r=-0.32, p=0.026$). This correlation suggests that when user's preference distributions are more concentrated, the recommended meals tend to perform better. This is not too suprising because the entropy of the preference distribution roughly reflects the degree of confidence the system has in the users' preferences, where the confidence is higher if the entropy is lower and vice versa. In Fig.~\ref{fig:entropy}, we show the evolution of the entropy value as the users are making more comparisons. The results demonstrate that the system becomes more confident about user's preferences as users provide more feedback.

\section{Discussion}
\label{sec:discussion}

In this section, we discuss the limitations of the current prototype and study and present real world scenarios where \textbf{Yum-me} and its sub-modules can be used.

\subsection{Limitations of the evaluations}
In evaluating the online learning framework, because there is no previous algorithm that can end-to-end solve our preference elicitation problem, the baselines are constructed by combining methods that intuitively fit \textit{user state update} and \textit{images selection} modules, respectively. This introduces potential biases in baseline selections. Additionally, in the end-to-end user testing, the participants' judgements of whether the food is \textit{Yummy} or \textit{No way} is potentially influenced by the image quality and the health concerns. These may be confounding factors in measuring users' preferences towards food items and can be eliminated by explicitly instructing the participants to not consider these factors. We leave further evaluations as future work.

\subsection{Limitations of Yum-me in recommending healthy meals}

The ultimate effectiveness of \textbf{Yum-me} in generating healthy meal suggestions is contingent on the appropriateness of the nutritional needs input by the user. In order to conduct such recommendations for people with different conditions, \textbf{Yum-me} could be used in the context of personal health coaches, nutritionists or coaching applications that provide reliable nutritional suggestions based on the user's age, weight, height, exercise and disease history. For instance, general nutritional recommendations can be calculated using online services built on the guidelines from National Institutes of Health, such as \textit{weight-success}\footnote{\url{http://www.weighing-success.com/NutritionalNeeds.html}} and \textit{active}\footnote{\url{http://www.active.com/fitness/calculators/nutrition}}. Also, although we have demonstrated the feasibility of building a personalized meal recommender catering to people's fine-grained food preference and nutritional needs,  the current prototype of \textbf{Yum-me} assumes a relatively simple strategy to rank the nutritional appropriateness, and is limited in terms of the available options for nutrition. Future work should investigate more sophisticated ranking approaches and incorporate options relevant to the specific  application context.

\subsection{Yum-me for real world dietary applications}
We envision that \textbf{Yum-me} has the potential to power many real-world dietary applications. For example, (1) \textbf{User onboarding}. Traditionally, food companies, e.g. Zipongo and Plated, address the cold start problem by asking each new user to answer a set of pre-defined questions, as shown in Section \ref{sec:yum-me user study}, and then recommend meals accordingly. \textbf{Yum-me} can enhance this process by eliciting user's fine-grained food preference and informing an accurate dietary profile. Service providers can customize \textbf{Yum-me} to serve their own businesses and products by using a specialized  backend food item database, and then use it as a step after the general questions. (2) \textbf{Nutritional assistants.} While visiting a doctor's office, patients are often asked to fill out standard questionnaires to indicate food preferences and restrictions. Patients' answers are then investigated by the professionals to come up with effective and personalized dietary suggestions. In such a scenario,  the recommendations made by \textbf{Yum-me} could provide a complementary channel for communicating the patient's fine-grained food preferences to the doctor to further tailor suggestions.

\subsection{FoodDist for a wide range of food image analysis tasks}

FoodDist provides a unified model to extract features from food images so that they are discriminative in the classification and clustering tasks, and its pairwise Euclidean distances are meaningful in reflecting similarities. The model is rather efficient ($<0.5$s/f on 8-core commodity processors) and can be ported to mobile devices with the publicly-available caffe-android-lib framework\footnote{\url{https://github.com/sh1r0/caffe-android-lib}}. 

In addition to enabling Yum-me, we released the FoodDist model to the community (\url{https://github.com/ylongqi/FoodDist}) so that it can be used to fuel other nutritional  applications. For the sake of space, we only briefly discuss two sample use cases below:

\begin{itemize}
    \item \textbf{Food/Meal recognition:} Given a set of labels, e.g., food categories, cuisines, and restaurants, the task of food and meal recognition could be approached by first extracting food image features from FoodDist and then training a linear classifier, e.g., logistic regression or SVM, to classify the food images that are beyond the categories given in the Food-101 dataset.
    \item \textbf{Nutrition Facts estimation:} With the emergence of large-scale food item or recipe databases, such as Yummly, the problem of nutritional fact estimation might be converted to a simple nearest-neighbor retrieval task: given a query image, we find its closest neighbor in the FoodDist based on Euclidean distance, and use that neighbor's nutritional information to estimate the nutrition facts of the query image~\cite{meyers2015im2calories}.
\end{itemize}

\section{Conclusion and Future work}
\label{sec:conclusion}

In this paper, we propose \textbf{Yum-me}, a novel nutrient-based meal recommender that makes meal recommendations catering to users' fine-grained food preferences and nutritional needs. We further present an online learning algorithm that is capable of efficiently learning food preference, and \textbf{FoodDist}, a best-of-its-kind unified food image analysis model. The user study and benchmarking results demonstrate the effectiveness of Yum-me and superior performance of FoodDist model. 

Looking forward, we envision that the idea of using visual similarity for preference elicitation may have implications to the following research areas. (1) \textbf{User-centric modeling}: the fine-grained food preference learned by Yum-me can be seen as a general dietary profile of each user and be projected to other domains to enable more dietary applications, such as suggesting proper meal plans for diabetes patients. Moreover, a personal dietary API can be built on top of this profile to enable sharing and improvementment across multiple dietary applications. (2) \textbf{Food image analysis API for deeper content understanding}: With the release of the FoodDist model and API, many dietary applications, in particular the ones that capture a large number of food images, might benefit from a deeper understanding of their image contents. For instance, food journaling applications could benefit from the automatic analysis of food images to summarize the day-to-day food intake or trigger timely reminders and suggestions when needed. (3) \textbf{Fine-grained preference elicitation leveraging visual interfaces.} The idea of eliciting users' fine-grained preference via visual interfaces is also applicable to other domains. The key insight here is that visual contents capture many subtle variations among objects that text or categorical data cannot capture; and the learned representations can be used as an effective medium to enable fine-grained preferences learning. For instance, the IoT, wearable, and mobile systems for entertainments, consumer products, and general content deliveries might leverage such an adaptive visual interface to design an onboarding process that learn users' preferences in a much shorter time and potentially provide a more pleasant user experience than traditional approaches.

\begin{acks}
We would like to thank the anonymous reviewers for their insightful
comments and thank Yin Cui, Fan Zhang, Tsung-Yi Lin, and Dr. Thorsten Joachims for discussion of machine learning algorithms. 
\end{acks}

\bibliographystyle{ACM-Reference-Format}
\bibliography{sample-bibliography}

\end{document}